\documentclass[final,3p,times,authoryear]{elsarticle}

\input{preamble}

\begin{document}

\begin{frontmatter}

\title{A topology optimisation framework to design test specimens for one‑shot identification or discovery of material models}

\author[1]{Saeid Ghouli}
\author[1]{Moritz Flaschel}
\author[2]{Siddhant Kumar}
\author[1]{Laura De Lorenzis\corref{cor}}\ead{ldelorenzis@ethz.ch}
\cortext[cor]{Corresponding author}
\address[1]{Department of Mechanical and Process Engineering, ETH Zürich, 8092 Zürich, Switzerland}
\address[2]{Department of Materials Science and Engineering, Delft University of Technology, 2628 CD Delft, The Netherlands}

\begin{abstract}
\label{abstract}
The increasing availability of full-field displacement data from imaging techniques in experimental mechanics is determining a gradual shift in the paradigm of material model calibration and discovery, from using several simple-geometry tests towards a few, or even one single test with complicated geometry. The feasibility of such a ``one-shot" calibration or discovery heavily relies upon the richness of the measured displacement data, i.e., their ability to probe the space of the state variables and the stress space (whereby the stresses depend on the constitutive law being sought) to an extent sufficient for an accurate and robust calibration or discovery process. The richness of the displacement data is in turn directly governed by the specimen geometry.
In this paper, we propose a density-based topology optimisation framework to optimally design the geometry of the target specimen for calibration of an anisotropic elastic material model. To this end, we perform automatic, high-resolution specimen design by maximising the robustness of the solution of the inverse problem, i.e., the identified material parameters, given noisy displacement measurements from digital image correlation. We discuss the choice of the cost function and the design of the topology optimisation framework, and we analyse a range of optimised topologies generated for the identification of isotropic and anisotropic elastic responses.
\end{abstract}


\end{frontmatter}

\section{Introduction}
\label{introduction}

Characterising the mechanical behaviour of materials typically requires multiple tests. 
For instance, fully characterising the linear elastic behaviour of a planar orthotropic material (such as a composite lamina) typically demands at least three separate experiments \citep{Tsai1965,Acosta-Flores2024}; these could be, for instance, a longitudinal tensile test (measuring longitudinal and transverse strains simultaneously), a transverse tensile test, and an in-plane shear test. Also, traditional mechanical testing involves measurements with contact sensors such as strain gauges or displacement transducers, providing local experimental data only. 
The advent of advanced full-field measurement techniques such as digital image correlation (\mbox{DIC}) and digital volume correlation (\mbox{DVC}) has revolutionised the field of material law characterisation by opening up the perspective of fully characterising material behaviour with one single test \citep{Guelon2009,Fu2020}.
In this new context, the optimal design of the target specimen is of utmost importance. A review paper by \cite{Pierron2021} denotes this paradigm as ``material testing 2.0". 

During the last decades, different groups of researchers have dealt with the specimen design problem suited for parameter identification via full-field measurements. 
As follows, we briefly highlight the major concepts and methodologies, focusing only on automatic (i.e., optimisation-based) approaches. 
Many studies have dealt with shape optimisation techniques to optimise the geometry of a pre-selected topology parametrised with a few variables. 
Different cost functions have been used for different classes of constitutive laws. In the context of orthotropic materials, \cite{Grediac1998} aimed at equal co-existence of different strain components in the test specimen. \cite{Pierron2007} targeted maintaining the balance between the sensitivities to noise for the identification of the different stiffness components. \cite{Gu2016a} accounted for the full identification process, thus optimising the \mbox{DIC} metrological parameters in addition to geometrical variables. The cost function was defined as the sum of systematic and random errors when calibrating the orthotropic stiffness components. Such a cost definition depends on the reference (i.e., ground-truth) material parameters. In the context of elastoplasticity, \cite{Souto2015} put forward the definition of a heterogeneity indicator (combining the standard deviation, range, maximum and average of different strain states) which could assess the suitability of a test specimen for sheet metal testing. The heterogeneity indicator was then employed in \cite{Souto2016} to design a butterfly-shaped specimen. \cite{Bertin2016} minimised the uncertainty in the parameter identification process by maximising the minimum eigenvalue of the Hessian matrix encountered in the finite element model updating (\mbox{FEMU}) approach. \cite{Chapelier2022} followed a similar approach and proposed a spline-based shape optimisation with special constraints to prevent excessive mesh distortion and load increase in a two-step shape optimisation, involving remeshing in between. \cite{Zhang2022} and \cite{Conde2023} employed the heterogeneity indicator proposed by \cite{Souto2015} in their cost functions, and confirmed that the initial configuration of the geometry has a strong influence on the final optimised design. More recently, \cite{Tung2024} devised a cost function quantifying the spread of data points in the deviatoric plane, while \cite{Ihuaenyi2024} utilised the information entropy concept to quantify the heterogeneity of the stress data points in the space of stress-triaxiality versus Lode angle parameter.

Another stream of research has employed topology optimisation techniques to automatically design optimal test specimens for constitutive parameter calibration facilitated by \mbox{DIC} measurements. 
Within the popular framework of density-based topology optimisation, the design variables are the virtual densities which designate the existence of void or material in each element of a finite element (\mbox{FE}) mesh \citep{Bendsøe1989}. Also in this stream, different cost formulations have been proposed. 
For orthotropic materials, \cite{Chamoin2020} followed a similar strategy to \cite{Bertin2016} to minimise the uncertainty in the parameter identification process. Owing to the computational cost associated with numerical gradient calculation, only coarse-resolution topologies including grey scales were generated. \cite{Almeida2020} employed the sum of fractions of the specimen surface undergoing certain principal stress states as a heterogeneity function, which was then used in a multi-objective optimisation setup together with the compliance to maintain sufficient stiffness in the designed specimen. The final designs had low resolutions, contained grey scales, and were difficult to realise and manufacture. The weights balancing the multi-objective optimisation problem were adjusted heuristically, while they were recognised to highly influence the output. \cite{Barroqueiro2020} adopted the concept of compliant mechanisms and proposed the ratio between the output and input displacements (of the mechanism with linear elastic response) as the cost function. Through a two-stage algorithm, the authors first produced a large number of heterogeneous designs, and then ranked them using a performance indicator (favouring equal co-existence of different deformation modes) while assuming an elastoplastic response. The best topology underwent a manual redesign to ensure smooth outer boundaries. \cite{Goncalves2023} improved the original framework proposed in \cite{Barroqueiro2020} by considering an additional, outer optimisation loop (leading to a two-level optimisation framework) which sought an optimal design domain configuration by changing parameters such as aspect ratio and material volume fraction. The authors performed numerous preliminary studies to define suitable bounds for the configurational parameters such that the heterogeneously designed specimen does not fail prematurely (i.e., prior to plasticity). The design strategy remained indirect and two-step while requiring expertise in selecting the parameter bounds and (dis-)approving the generated designs.

The above short review reveals some gaps in the current state of the art. 
Shape optimisation approaches, while leading to much faster computations, require significant expertise in material testing for the initial selection of a suitable topology.
On the other hand, when employing topology optimisation, 
the need to reduce the computational cost 
(due to the exploding number of design variables and possibly the employment of finite differences for sensitivity calculation) 
often results in low-resolution or grey-scale-diluted topologies.
Two-step processes require manual intervention.
Also, it seems that not much attention has been paid so far to global convergence (i.e., mesh independence) of the optimised topologies
to ensure their consistency 
as the \mbox{FE} discretisation of the design domain is refined. This property indicates the appropriateness of the cost definition for driving a stable optimisation process until convergence, and confirms resilience against small perturbations of the initialisation. 
Lastly, in the case of multi-objective optimisation (e.g. to improve manufacturability), the determination of weight factors balancing the different cost terms has been found to greatly impact the optimisation outcome. Yet, a non-systematic heuristic approach has been adopted. 

In this work, we propose a new topology optimisation framework for robust identification of constitutive parameters. We employ the density-based topology optimisation approach, 
and choose a high resolution to allow for great flexibility in the specimen design. 
Through filtering techniques, we provide binary, black-and-white designs. The cost function for the topology optimisation problem is defined based on our recently proposed material model discovery approach denoted as EUCLID \citep{Flaschel2021a}, is
independent of the ground-truth material parameters and 
targets the stability of the identification problem equations. 
We compute the sensitivities analytically and in a vectorised fashion. Moreover, we investigate and prove the global convergence of the optimised topologies.
To account for manufacturability and avoid possible weak regions in the specimen design, we employ the concept of robust topology optimisation which automates the redesigning process by a three-field projection filtering. We present the resulting optimised topologies for a range of linear elastic isotropic and orthotropic materials and assess their performance in terms of error in parameter identification. 
This paper is organised as follows. In \cref{parameter_identification} we formulate the constitutive parameter identification framework within anisotropic linear elasticity, which provides the basis for our topology optimisation framework in \cref{topopt}. \cref{results_discuss} presents the main results and also discusses the concepts of global and local convergence.
We conclude the paper in \cref{conclusions}.

\section{Parameter identification framework}
\label{parameter_identification}

\begin{figure}[tb]
    \centering
    \includegraphics[width=0.3\linewidth]{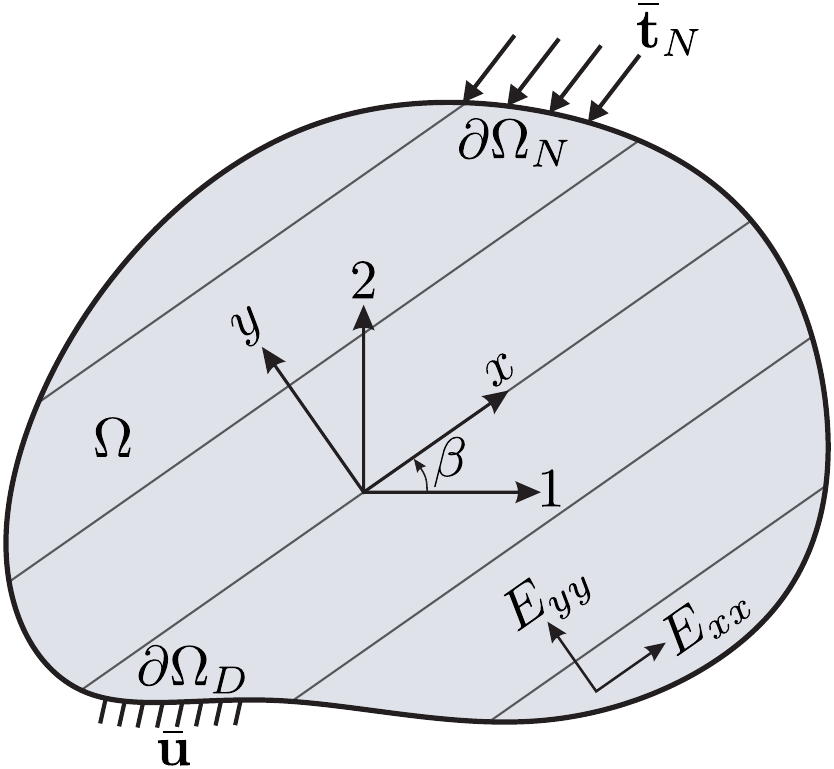}
    \caption[]{The boundary value problem of plane deformation for the case of an orthotropic elastic medium. The $x$-$y$ coordinate system aligns with the anisotropy orientation $\beta$ indicated by hatches and measured counterclockwise from the reference coordinates $1$-$2$.}
    \label{fig_01}
\end{figure}

In this section, we establish the system of equations leading to the unknown constitutive parameters. 
The input data consist of a (noisy) deformation field captured by \mbox{DIC} from the surface of a target specimen. 
We consider a deformable solid occupying the 2D domain $\Omega$ with boundary $\partial \Omega$ (\cref{fig_01}). Assuming small deformations and neglecting body forces, the principle of virtual work reads
\begin{equation} \label{eq_cost01}
     \int_\Omega \ten{\sigma}:\nabla{ \left(\delta \ten{u}\right)} \,  dA - \int_{\partial\Omega_N} \ten{\bar{t}}_N \cdot \delta \ten{u} \, ds \overset{!}{=} 0 \, ,
\end{equation}
\noindent
where $\ten{u}$ is the displacement field satisfying Dirichlet boundary conditions (i.e., $ \ten{u}=\bar{\ten{u}}$ on $\partial \Omega_D$), the virtual displacement $\delta \ten{u}$ is sufficiently regular and satisfies homogeneous Dirichlet boundary conditions (i.e., $\delta \ten{u}=\mathbf{0}$ on $\partial \Omega_D$) but is otherwise arbitrary, $\ten{\sigma}$ is the Cauchy stress tensor, and $\ten{\bar{t}}_N$ is the traction applied on the Neumann portion of the boundary $\partial\Omega_N$. 
The FE approximation of \cref{eq_cost01} defined over the domain $\Omega^\mathcal{h}$ discretised into $n_e$ elements (each with domain $\Omega_e$) reads
\begin{equation} \label{eq_cost02} \sum_{e=1}^{n_e}\int_{\Omega_e} \ten{\sigma}_e^\mathcal{h}:\nabla{\left(\delta \ten{u}_e^\mathcal{h}\right)} \,  dA -\sum_{e:\partial\Omega_e \subset \partial\Omega_N^\mathcal{h}} \int_{\partial\Omega_e} \ten{\bar{t}}_{N_e} \cdot \delta \ten{u}_e^\mathcal{h} \, ds \overset{!}{=} 0 \, ,
\end{equation}
\noindent
with superscript $\mathcal{h}$ and subscript $e$ denoting FE discretisation and element-related quantities, respectively.
For discretisation we 
adopt standard isoparametric FEs with
linear shape functions $N^a{\left(\ten{\xi}\right)}$, i.e.
\begin{equation} \label{eq_cost03}
 \ten{u}_e^\mathcal{h}\left(\ten{\xi}\right) = \sum_{a=1}^{n_{n,e}} N^a{\left(\ten{\xi}\right)} \ten{u}_e^a \, ,   \qquad
 \delta\ten{u}_e^\mathcal{h}\left(\ten{\xi}\right) =  \sum_{a=1}^{n_{n,e}} N^a{\left(\ten{\xi}\right)} \delta\ten{u}_e^a \, ,  \qquad
  \ten{x}_e^\mathcal{h}\left(\ten{\xi}\right) =  \sum_{a=1}^{n_{n,e}} N^a{\left(\ten{\xi}\right)} \ten{x}_e^a \, ,
\end{equation}
\noindent
where superscript $a$ refers to nodal quantities, 
$n_{n,e}$ is the number of nodes of element $e$, and 
parent space coordinates $\ten{\xi}$ are related to $\ten{x}$ through the  Jacobian $\ten{J}_e = \partial \ten{x}_e^\mathcal{h}/\partial \ten{\xi}$. 
We collect the nodal displacements in ${\ten{U}_e \in \mathbb{R}^{\abs{\mathcal{D}_e}}}$, where $\mathcal{D}_e = \{(a,i): a=1,\ldots,n_{n,e}; i=1,2\}$ designates the collection of all the degrees of freedom (DOFs) of element $e$, and accordingly arrange the shape functions $N^a$ in the matrix ${\ten{N} \in \mathbb{R}^{\abs{\mathcal{D}_e} \times 2}}$ and their gradients ${\nabla_{\ten{x}} \ten{N} = \ten{J}_e^{-T} \nabla_{\ten{\xi}} \ten{N}}$ as the transpose of ${\ten{B}_e \in \mathbb{R}^{3 \times \abs{\mathcal{D}_e}}}$. Rewriting \cref{eq_cost02} with integration over the parent element domain $\Omega_\smallsquare$ gives
\begin{equation} \label{eq_cost05}
    \delta \ten{U}^T \left(\sum_{e=1}^{n_e} \underbrace{\int_{\Omega_\smallsquare} \ten{B}_e^T{\!\left(\ten{\xi} \right)} \hat{\ten{\sigma}}_e^\mathcal{h}{\! \left(\ten{\xi} \right)} \det{\!\left(\ten{J}_e{\left(\ten{\xi} \right)} \right)} \, dA_\smallsquare}_{\ten{F}_{int,e}} -\sum_{e:\partial\Omega_e \subset \partial\Omega_N^\mathcal{h}} \underbrace{\int_{\partial\Omega_\smallsquare}  \ten{N}^T{\!\left(\ten{\xi}\right)} \ten{\bar{t}}_{N_e} j_e^s \,  ds_\smallsquare}_{\ten{F}_{ext,e}} \right) \overset{!}{=} 0 \, ,
\end{equation}
\noindent
where we use Voigt notation for the stress $\hat{\ten{\sigma}}_e^\mathcal{h}=\left[\sigma_{11},\sigma_{22},\tau_{12} \right]^T$, $j_e^s$ is the length Jacobian for the boundary mapping between parent and physical element, and we introduce the assembly operation ${\ten{U}=\bigcup_{e=1}^{n_e} \ten{U}_e}$ with $\ten{U} \in \mathbb{R}^{\abs{\mathcal{D}}}$, $\mathcal{D}$ as the set of all nodal DOFs, i.e., $\mathcal{D}=\{(a,i): a=1,\ldots,n_n; i=1,2\}$, and $n_n$ as the total number of nodes. In \cref{eq_cost05}, the two integrals represent the element internal and external force vectors, $\ten{F}_{int,e}, \ten{F}_{ext,e} \in \mathbb{R}^{\abs{\mathcal{D}_e}}$, respectively.
Assuming now anisotropic linear elasticity, we can express the Cauchy stress vector $\hat{\ten{\sigma}}_e^\mathcal{h}$ as a function of the strain vector, also expressed in Voigt notation as $\hat{\ten{\varepsilon}}_e^\mathcal{h} = \left[\varepsilon_{11},\varepsilon_{22},\gamma_{12} \right]^T = {\ten{B}_e \ten{U}_e}$, in two equivalent formats as follows

\begin{equation}
    \hat{\ten{\sigma}}_e^\mathcal{h} =   
    \tilde{\rho}_e^\mathrm{phys} 
    \begin{bmatrix} D_{11} & D_{12} & D_{16} \\
                           & D_{22} & D_{26} \\
                    sym.   &        & D_{66} \end{bmatrix}
    \begin{Bmatrix} \varepsilon_{11} \\ \varepsilon_{22} \\ \gamma_{12} \end{Bmatrix} \, =   
    \tilde{\rho}_e^\mathrm{phys}
    \underbrace{
    \begin{bmatrix} \varepsilon_{11} & \varepsilon_{22} & \gamma_{12} & 0 & 0 & 0 \\
                    0 & \varepsilon_{11} & 0 & \varepsilon_{22} & \gamma_{12} & 0 \\
                    0 & 0 & \varepsilon_{11} & 0 & \varepsilon_{22} & \gamma_{12} \end{bmatrix}}_{\tilde{\ten{\varepsilon}}_e^\mathcal{h}}
    \underbrace{
    \begin{Bmatrix} D_{11} \\ D_{12} \\ D_{16} \\ D_{22} \\ D_{26} \\ D_{66}  \end{Bmatrix}}_{\ten{\theta}} =
    \tilde{\rho}_e^\mathrm{phys} \tilde{\ten{\varepsilon}}_e^\mathcal{h} \ten{\theta}  \, . \label{eq_cost06b}
\end{equation}
\noindent
Here, the factor $\tilde{\rho}_e^\mathrm{phys}$ is the modified physical density of element $e$ in the context of topology optimisation, to be better specified later. In the second equality of \cref{eq_cost06b}, the strain components are cast in the matrix ${\tilde{\ten{\varepsilon}}_e^\mathcal{h} \in \mathbb{R}^{3 \times n_f}}$ and the elements of the elasticity matrix $\mathbf{D}$ are gathered in the vector of the unknown material parameters ${\ten{\theta} \in \mathbb{R}^{n_f}}$, where $n_f=6$. Employing \cref{eq_cost06b} in the definition of the element internal force vector from \cref{eq_cost05} gives
\begin{equation} \label{eq_cost07}
    \ten{F}_{int,e} = \tilde{\rho}_e^\mathrm{phys} \underbrace{\left( \int_{\Omega_\smallsquare} \ten{B}_e^T{\! \left(\ten{\xi} \right)} \tilde{\ten{\varepsilon}}_e^\mathcal{h}{\!\left(\ten{\xi}\right)} \det{\!\left(\ten{J}_e {\left(\ten{\xi} \right)} \right)} \, dA_\smallsquare \right)}_{\ten{A}_e} \ten{\theta} = \tilde{\rho}_e^\mathrm{phys}  \ten{A}_e \, \ten{\theta} \, ,
\end{equation}
\noindent
where we have assumed spatial invariance of the material parameters $\ten{\theta}$ and designated the element integral by ${\ten{A}_e \in \mathbb{R}^{\abs{\mathcal{D}_e} \times n_f}}$. Using $n_\mathrm{gp}$ Gauss quadrature points per parametric direction with coordinates $\left(\xi_i,\eta_j \right)$ ($i,j=1,\ldots,n_\mathrm{gp}$) and corresponding weights $w_i,w_j$, $\ten{A}_e$ is computed as
\begin{equation} \label{eq_cost08}
    \ten{A}_e{\left(\hat{\ten{\varepsilon}}_e^\mathcal{h}{\!\left(\ten{U}_e\right)} \right)} = \int_{\Omega_\smallsquare} \ten{B}_e^T{\! \left(\ten{\xi} \right)} \tilde{\ten{\varepsilon}}_e^\mathcal{h}{\! \left(\ten{\xi}\right)}  \det{\!\left(\ten{J}_e{\left(\ten{\xi} \right)} \right)} \, dA_\smallsquare \approx \sum_{i=1}^{n_\mathrm{gp}} {\sum_{j=1}^{n_\mathrm{gp}} {\ten{B}_e^T{\! \left(\xi_i,\eta_j \right)} \tilde{\ten{\varepsilon}}_e^\mathcal{h}{\! \left(\xi_i,\eta_j \right)}  \det{\!\left(\ten{J}_e{ \left(\xi_i,\eta_j \right)} \right)} \times w_i \times w_j }} \, .
\end{equation}
\noindent
Row-wise assembly over $\ten{F}_{int,e}$ from \cref{eq_cost07} then gives the global matrix
\begin{equation} \label{eq_cost09}
    \ten{A}_\mathrm{glob}{\left({\ten{\rho}}^\mathrm{phys},\ten{U}\right)} = \bigcup_{e=1}^{n_e} {\tilde{\rho}_e^\mathrm{phys}{\left({\rho}_e^\mathrm{phys}\right)} \, \ten{A}_e{\left(\ten{U}_e \right)}}  \, .
\end{equation}
\noindent
The dependence of ${\ten{A}_{\mathrm{glob}} \in \mathbb{R}^{\abs{\mathcal{D}} \times n_f}}$ on the vector of element physical densities ${\ten{\rho}}^\mathrm{phys} \in \mathbb{R}^{n_e}$ and on $\ten{U}$ will be exploited in \cref{problem}. Both $\ten{A}_{\mathrm{glob}}$ and $\ten{A}_e$ have unit of length.

The set of all DOFs $\mathcal{D}$ consists of two subsets, namely the free and fixed (i.e., Dirichlet boundary) DOFs: ${\mathcal{D}^\mathrm{free} \! \subseteq \! \mathcal{D}}$ and ${\mathcal{D}^\mathrm{fix} = \mathcal{D} \! \setminus \! \mathcal{D}^\mathrm{free}}$. 
The equilibrium equation for the free DOFs
can be obtained from \cref{eq_cost05} using \cref{eq_cost09}:
\begin{equation} \label{eq_cost10}
    \ten{A}_\mathrm{glob}^\mathrm{free} \, \ten{\theta} \overset{!}{=} \ten{b}_\mathrm{glob}^\mathrm{free} \, ,
\end{equation}
\noindent
where ${\ten{A}_\mathrm{glob}^\mathrm{free} \in \mathbb{R}^{\abs{\mathcal{D}^\mathrm{free}} \times n_f}}$ and ${\ten{b}_\mathrm{glob}^\mathrm{free} \in \mathbb{R}^{\abs{\mathcal{D}^\mathrm{free}}}}$. Here, we note that ${\ten{b}_\mathrm{glob}^\mathrm{free}}$ contains the integrated global traction forces over the Neumann boundary (in a displacement-controlled setting, ${\ten{b}_\mathrm{glob}^\mathrm{free}=\ten{0}}$).

For the fixed DOFs, the internal force from the material must balance out with the reaction force. Practically, the reaction force is not known at every fixed DOF; rather, it is only the sum of the reaction forces $\bar{R}^s$ that can be measured experimentally at certain subsets of the fixed DOFs denoted as $\mathcal{D}^{\mathrm{fix},s}$ (${s=1,\ldots,n_s}$; ${\cup_{s=1}^{n_s} {\mathcal{D}^{\mathrm{fix},s}} \subseteq \mathcal{D}^\mathrm{fix}}$; ${\mathcal{D}^{\mathrm{fix},s} \! \cap \!  \mathcal{D}^{\mathrm{fix},t} = \emptyset \text{ for } s \neq t}$). The force balance equation for each subset $s$ can thus be written as
\begin{equation} \label{eq_cost11}
    \sum_{\mathcal{D} \subset \mathcal{D}^{\mathrm{fix},s}} {\ten{A}_\mathrm{glob}^{\mathrm{fix},s}} \, \ten{\theta} = \bar{R}^s \, .
\end{equation}
\noindent
Note that each subset contains DOFs corresponding to displacement components only in one direction. The collection of these individual equations for all $n_s$ subsets of the fixed DOFs will then provide us with
\begin{equation} \label{eq_cost12}
    \ten{A}_\mathrm{glob}^\mathrm{fix} \, \ten{\theta} \overset{!}{=} \ten{b}_\mathrm{glob}^\mathrm{fix} \, ,
\end{equation}
\noindent
wherein ${\ten{A}_\mathrm{glob}^\mathrm{fix} \in \mathbb{R}^{n_s \times n_f}}$ and ${\ten{b}_\mathrm{glob}^\mathrm{fix} \in \mathbb{R}^{n_s}}$. Note that, in a uniaxial testing setup, reaction forces are known only at the top and bottom fixed edges of the specimen (where they are equal and opposite), therefore $n_s = 2$.

\Cref{eq_cost10,eq_cost12} are systems of linear equations which must be satisfied simultaneously by the solution $\ten{\theta}$. Therefore, we combine the two sets of equations by concatenating them vertically and defining $\ten{A} \in \mathbb{R}^{(\abs{\mathcal{D}^\mathrm{free}} + n_s) \times n_f}$ and $\ten{b} \in \mathbb{R}^{\abs{\mathcal{D}^\mathrm{free}} + n_s}$ as 
\begin{equation} \label{eq_cost13}
    \ten{A} = \sqrt{\lambda_q} \, [\ten{A}_\mathrm{glob}^\mathrm{free} \, ; \sqrt{\lambda_r} \, \ten{A}_\mathrm{glob}^\mathrm{fix}]; \quad  \ten{b} = \sqrt{\lambda_q} \, [\ten{b}_\mathrm{glob}^\mathrm{free} ; \sqrt{\lambda_r} \, \ten{b}_\mathrm{glob}^\mathrm{fix}] \, .
\end{equation}
\noindent
Factor $\sqrt{\lambda_r}$ is used to balance the contribution of the fixed and free DOFs in the system of equations, whereas factor $\sqrt{\lambda_q}$ accounts for mesh convergence, 
 see \ref{app_weighting_factor} for details. The final equilibrium equation reads
\begin{equation} \label{eq_cost14}
    \ten{A} \ten{\theta} \overset{!}{=} \ten{b} \, .
\end{equation}
\noindent
Note that $\ten{A}$ is a purely kinematic quantity, with dimension of a length, which encodes information only on the geometry, loading configuration and boundary conditions, while the material properties are contained in $\ten{\theta}$.
Since $\abs{\mathcal{D}^\mathrm{free}} + n_s \gg n_f$, \cref{eq_cost14} is over-determined and can be solved in a least squares sense by
\begin{equation} \label{eq_cost17}
    \ten{A}^\mathrm{eqb} \ten{\theta} \overset{!}{=} \ten{b}^\mathrm{eqb} \, ,
\end{equation}
\noindent
with
\begin{subequations} \label{eq_cost16}
\begin{gather}
    \ten{A}^\mathrm{eqb} = \ten{A}^T \ten{A} = \lambda_q \left( \left(\ten{A}_\mathrm{glob}^{{\mathrm{free}}^T} \ten{A}_\mathrm{glob}^\mathrm{free} \right) + \lambda_r \left(\ten{A}_\mathrm{glob}^{{\mathrm{fix}}^T} \ten{A}_\mathrm{glob}^\mathrm{fix} \right) \right) \, \in \mathbb{R}^{n_f \times n_f} \, , \label{eq_cost16a} \\
    \ten{b}^\mathrm{eqb} = \ten{A}^T \ten{b} = \lambda_q \left( \left(\ten{A}_\mathrm{glob}^{{\mathrm{free}}^T} \ten{b}_\mathrm{glob}^\mathrm{free} \right) + \lambda_r \left(\ten{A}_\mathrm{glob}^{{\mathrm{fix}}^T} \ten{b}_\mathrm{glob}^\mathrm{fix} \right) \right) \in \mathbb{R}^{n_f} \, , \label{eq_cost16b}
\end{gather}
\end{subequations}
where the symmetric positive definite matrix $\ten{A}^\mathrm{eqb}$ and the right-hand side $\ten{b}^\mathrm{eqb}$ have dimensions of a square length and a force, respectively.
A well-known result in numerical analysis \citep{Ascher2011} gives the following estimate
\begin{equation} \label{eq_cost24}
     \dfrac{{\norm{\delta\ten{\theta}}_2}}{{\norm{\ten{\theta}_\mathrm{}}}_2} \leq \kappa_2{\left(\ten{A}^\mathrm{eqb}\right)} \dfrac{{\norm{\delta\ten{b}^\mathrm{eqb}}}_2}{{\norm{\ten{b}^\mathrm{eqb}}}_2} \, ,
\end{equation}
\noindent
where $\delta\ten{b}^\mathrm{eqb}$ and $\delta\ten{\theta}$ are the perturbations (i.e., noise) in the problem and the induced error in the solution, respectively, and  
\begin{equation} \label{eq_cost23}
     \kappa_2{\left(\ten{A}^\mathrm{eqb}\right)} = {\norm{{\ten{A}^\mathrm{eqb}}^{\phantom{l\!}}}}_2 \, {\norm{{\ten{A}^\mathrm{eqb}}^{-1}}}_2 
\end{equation}
is the $2$-norm condition number of $\ten{A}^\mathrm{eqb}$. 
This inequality indicates that the $\mathcal{l}_2$-error magnification from the problem to the solution is bounded by $\kappa_2{\left(\ten{A}^\mathrm{eqb}\right)}$. In other words, it is the condition number which governs the stability of the solution $\ten{\theta}$ upon perturbation in the problem. 
Hence, in order to minimise the uncertainty/maximise the robustness of the constitutive law identification against the noise in the deformation data, $\kappa_2(\ten{A}^\mathrm{eqb})$ should be minimised.
As discussed earlier, $\ten{A}^\mathrm{eqb}$ is a purely kinematic quantity which depends, among others, on the geometry of the domain, i.e., the test specimen used for material testing. This fact inspires the possibility of defining an appropriate cost function targeting the minimisation of $\kappa_2(\ten{A}^\mathrm{eqb})$ and utilising it in an optimisation framework to optimally design the test specimen. 

\section{Topology optimisation framework}
\label{topopt}

Topology optimisation seeks to find the best material layout in a design domain in order to minimise a certain cost function given a number of design constraints. Among different well-established approaches \citep{Sigmund2013}, we employ the density-based topology optimisation approach \citep{Bendsøe1989}, where the design variables are the virtual densities assigned to the FEs in the discretised domain. The virtual densities can acquire values between zero and one, representing the (non-)existence of material. In the following, we describe the design domain and define our choice of the cost function.

\subsection{Setup}
\label{setup}
\begin{figure}[tb]
\centering
    \begin{subfigure}{0.15\textwidth}
        \centering
        \includegraphics[width=\linewidth]{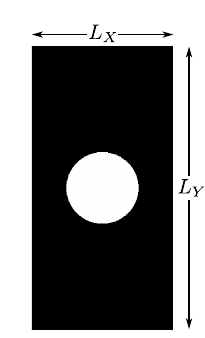} \caption{} \label{fig_02_a}
    \end{subfigure}
    \begin{subfigure}{0.15\textwidth}
        \centering
        \includegraphics[width=\linewidth]{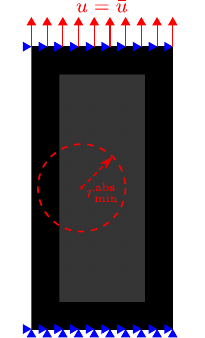} \caption{} \label{fig_02_b}
    \end{subfigure}
    \begin{subfigure}{0.15\textwidth}
        \centering
        \includegraphics[width=\linewidth]{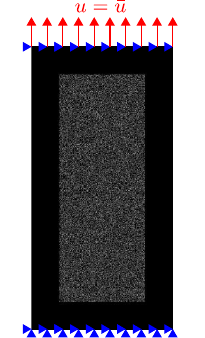} \caption{} \label{fig_02_c}
    \end{subfigure}
    \begin{subfigure}{0.15\textwidth}
        \centering
        \includegraphics[width=\linewidth]{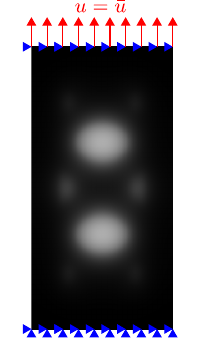} \caption{} \label{fig_02_d}
    \end{subfigure}
    \begin{subfigure}{0.115\textwidth}
        \centering
        \includegraphics[width=\linewidth]{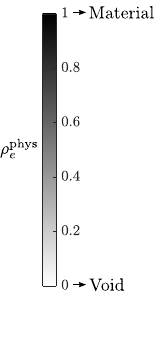} 
    \end{subfigure}
    \caption[]{(a) Designation of 0 (void) and 1 (material) densities as well as domain dimensions; (b) even, (c) random, and (d) interim-valued initialisations of the domain; Dirichlet boundary conditions resembling the uniaxial tensile test are schematised. The absolute filtering radius $r_\mathrm{min}^\mathrm{abs}$ is also visualised in figure (b).}
    \label{fig_02}
\end{figure}

\cref{fig_02_b} shows the rectangular domain, with dimensions $L_X$ and $L_y$, and the Dirichlet boundary conditions corresponding to a displacement-controlled tensile test with imposed displacement $\bar{u}=0.005 \times L_Y$ uniformly applied on the top edge. Note that in principle it would be possible to consider multiple load cases and optimise the topology for all the considered load cases. Herein, we stick to the uniaxial loading setup which is very common in the experimental practice. The black frame around the design domain is the so-called \textit{passive} region whose density values are kept fixed at $1$ with no changes allowed, while the inner grey area is open to change and optimisation. We always account for this passive frame around the design domain to ensure the preservation of the gripping areas at the top and bottom of the specimen (where the testing machine holds the specimen, i.e., Dirichlet boundaries), and to prevent the formation of holes and narrow strips of material in the proximity of the edges (which can be prone to premature failure upon loading and are often deemed inapplicable for \mbox{DIC} measurements). While the use of a fixed outer frame introduces small limitations to the design space, our approach remains highly versatile, as we will demonstrate below.

\subsection{Choice of the cost function}
\label{problem}

In order to mitigate the error in the calibration of the constitutive parameters and enhance their stability in the presence of noise, we aim to utilise topology optimisation to automatically design the topology of the test specimen so as to minimise the $2$-norm condition number $\kappa_2(\ten{A}^\mathrm{eqb})$ in \cref{eq_cost23}. 
A more general candidate is the $p$-norm condition number $\kappa_p$ defined as 
\begin{equation} \label{eq_topopt01}
    \kappa_p{\left(\ten{A}^\mathrm{eqb}\right)} = {\norm{{\ten{A}^\mathrm{eqb}}^{\phantom{l\!}}}}_p \, {\norm{{\ten{A}^\mathrm{eqb}}^{-1}}}_p  \, ,
\end{equation}
\noindent
where
\begin{equation} \label{eq_topopt02}
    {\norm{\ten{A}^\mathrm{eqb}}}_p = \left(\sum_{i=1}^{n_f}{ \sum_{j=1}^{n_f} {{\abs{A_{ij}^\mathrm{eqb}}}^p}} \right)^{\frac{1}{p}} 
    \, ,  
\end{equation}
\noindent
\noindent
with $p>2$, is the $p$-norm of matrix $\ten{A}^\mathrm{eqb}$. 
The bounds reported in \cref{app_p_norm_cond} show that the minimisation of $\kappa_p$ induces the minimisation of $\kappa_2$.
Hence, a candidate cost function is
\begin{equation} \label{eq_topopt04}
    \mathrm{cost}_{\mathrm{alt},1}{\left({\ten{\rho}}^\mathrm{phys},\ten{U}\right)} = \dfrac{\kappa_p{\left(\ten{A}^\mathrm{eqb}{\left({\ten{\rho}}^\mathrm{phys},\ten{U}\right)}\right)}^p}{\kappa_p{\left(\ten{A}_{@ \mathrm{init.}}^\mathrm{eqb}\right)}^p} \, ,
\end{equation}
\noindent
which is continuously differentiable away from zero. Based on our numerical experiments reported later, 
$p\geq8$ 
leads to a stable topology optimisation process with gradual topology updates, whereas 
$p<8$ behaves rather unstably and causes abrupt topology changes. Division by the initial value $\kappa_p{\left(\ten{A}_{@ \mathrm{init.}}^\mathrm{eqb}\right)}^p$ 
normalises the cost function and brings more stability to the optimisation process with more gradual topology updates. An advantage of optimising a condition number (be it $\kappa_2$, $\kappa_p$ with $p\neq2$ or condition numbers in other norms) is the insensitivity to the strain level induced in the specimen; this is owing to the fact that the condition number comprises the product of the norms of the matrix and the matrix inverse, hence cancelling out the strain magnitude. This feature enables the specimen design not to solely rely on high-strain concentration scenarios, but rather promote milder strain variations across the specimen and yet improve the identifiability of the material parameters.
Inspired by \cite{Chamoin2020} and \cite{Zhang2022}, we evaluate another possible cost function, namely, 
$1/\det{\!(\ten{A}^\mathrm{eqb})}$, 
which is also continuously differentiable:
\begin{equation} \label{eq_topopt05}
    \mathrm{cost}{\left({\ten{\rho}}^\mathrm{phys},\ten{U}\right)} = \dfrac{\det{\!\left(\ten{A}_{@ \mathrm{init.}}^\mathrm{eqb}\right)} }{\det{\!\left(\ten{A}^\mathrm{eqb}{\left({\ten{\rho}}^\mathrm{phys},\ten{U}\right)}\right)}} \,,
\end{equation}
\noindent
where the initial value, i.e., $\det{\!\left(\ten{A}_{@ \mathrm{init.}}^\mathrm{eqb}\right)}$, non-dimensionalises the cost and alleviates the dependence on the strain level (i.e., the imposed external displacement).
The minimisation of this cost
drives the matrix away from singularity, 
but a priori does not necessarily result in the minimisation of the condition number \citep{Ascher2011}. In our numerical experiments, adopting these and a few more options for the cost function (see also \ref{app_other_cost}), we found $1/\det{\!(\ten{A}^\mathrm{eqb})}$ to give the best performance, produce topologies with smoother boundaries, and drive the topology to a low condition number. In \cref{convergence} we provide some comparisons between results obtained with \cref{eq_topopt04,eq_topopt05}.

\subsection{Density-based topology optimisation}
\label{density_based_topology_optimisation}
We employ density-based topology optimisation
\citep{Bendsøe1989,Andreassen2011}, 
where the design variables are the 
element densities $\rho_e$ listed in the vector $\ten{\rho} \in \mathbb{R}^{n_e}$. Here, $\rho_e = 0$ and $\rho_e = 1$ respectively denote no material, i.e., void, and full material, colour-coded as white and black (see \cref{fig_02_a}). The intermediate densities, i.e. $0 < \rho_e < 1$, correspond to different shades of grey. In \cref{filter} we elaborate upon the transformation of $\rho_e$ to $\rho_e^\mathrm{phys}$ through \textit{filtering} which maps intermediate densities to either $0$ or $1$ to culminate in a physically realisable topology. With the mentioned parametrisation, the topology optimisation problem can be interpreted as finding the optimal material density distribution over a given design domain leading to cost minimisation. 
Hence, we define our optimisation problem as follows:
\begin{subequations} \label{eq_topopt06}
\begin{gather}
    \underset{\ten{\rho}}{\mathrm{min}}\!: \, \mathrm{cost}{\left({\ten{\rho}}^\mathrm{phys},\ten{U}\right)} \quad \text{or} \quad \mathrm{cost}_{\mathrm{alt},i}{\left({\ten{\rho}}^\mathrm{phys},\ten{U}\right)} , \, (i=1\ldots3) \, , \label{eq_topopt06a} \\
    s.t.: \left\{ \begin{array}{{l}}
        {\ten{K} \ten{U} = \ten{F}} \, , \\
        {\dfrac{1}{n_e}\sum_{e=1}^{n_e}{\rho_e^\mathrm{phys}} = V_m} \, , \\ 
        {0 \leq \ten{\rho} \leq 1} \, .
        \end{array} \right. \label{eq_topopt06b}
\end{gather}
\end{subequations}
\noindent
The problem at hand is a constrained optimisation problem, where the constraints are expressed in \cref{eq_topopt06b}. The first equality constraint is the state (i.e. equilibrium) equation (here for linear elasticity), with $\ten{K} \in \mathbb{R}^{\abs{\mathcal{D}} \times \abs{\mathcal{D}}}$ as the global stiffness matrix and $\ten{F} \in \mathbb{R}^{\abs{\mathcal{D}}}$ as the global external force vector (zero in displacement control). The second equality constraint fixes the average of the total amount of material (used in the topology design) to be equal to the material volume fraction $V_m$. This volume constraint is well-known for stabilising the optimisation process and facilitating numerical convergence \citep{Sigmund2013}. The last constraint is an inequality (or a bound) constraint which allows for the smooth variation of the design variables only in the range between $0$ and $1$.

In each iteration of the optimisation process, the nodal displacements $\ten{U}$ are first solved for from the equilibrium equation through FE analysis with the global stiffness matrix
\begin{equation} \label{eq_topopt07}
    \ten{K}{\left({\ten{\rho}}^\mathrm{phys},\ten{\theta}\right)} = \bigcup_{e=1}^{n_e} {\tilde{\rho}_e^\mathrm{phys}{\left(\rho_e^\mathrm{phys}\right)} \, \ten{K}_e{\left(\ten{\theta}\right)}} \, ,
\end{equation}
\noindent
where the density assigned to each element directly influences the significance of the contribution of that element stiffness $\ten{K}_e{(\ten{\theta})} \in \mathbb{R}^{\abs{\mathcal{D}_e} \times \abs{\mathcal{D}_e}}$ to the global stiffness of the domain. %We note that 
The stiffness matrix depends on the material parameters $\ten{\theta}$ which are the unknowns of the identification problem. Later in \cref{results_discuss} we discuss how the choice of these parameters affects the optimised topologies. The modified physical density in \cref{eq_topopt07} is 
given by
\begin{equation} \label{eq_topopt08}
     \tilde{\rho}_e^\mathrm{phys} = \rho_\mathrm{min} + \left(\rho_e^\mathrm{phys}\right)^{3} \left(1-\rho_\mathrm{min} \right) \, .
\end{equation}
\noindent
It is used to avoid singularity in the global stiffness matrix when $\rho_e^\mathrm{phys} = 0$, by adopting a threshold value $\rho_\mathrm{min}$ (here set to $10^{-9}$). Moreover, the power of $3$ is considered as a ``magic" number which helps drive the grey scales towards $0$ or $1$ and maintain numerical convergence \citep{Sigmund2013}. This number has been confirmed to ensure the physical realisability of intermediate densities \citep{Bendsøe1999}, with which \cite{Amstutz2011} verified the equality of density gradients and topological derivatives for elasticity. Regarding the discretisation, we employ bilinear quadrilateral elements with 4 Gauss points under plane-stress conditions, maintain equal size for all elements%(i.e., $v_e=1$)
, and keep a fixed uniform mesh throughout the optimisation process.
Topology optimisation without regularisation is an ill-posed problem which can lead to checkerboard patterns \citep{Andreassen2011}. %To rectify this, and to ensure the existence of solutions to the topology optimisation problem, 
For this reason, \textit{filtering} techniques have been developed. Filtering serves as a regularisation scheme, transforming the sharp, binary, $0$-$1$ topology (represented by $\ten{\rho}$) to a smeared, grey-scale topology (represented by $\ten{\rho}^\mathrm{avg}$) and is necessary to achieve convergence to a local minimum. In \ref{filter} we summarise explicit and implicit filtering approaches to transform $\ten{\rho}$ to $\ten{\rho}^\mathrm{avg}$. Therein, we justify our choice of an implicit \mbox{PDE} filter defined in \cref{eq_topopt13}. The grey scale densities $\ten{\rho}^\mathrm{avg}$ are undesirable in the output topology since they cannot be manufactured in practice. To convert $\ten{\rho}^\mathrm{avg}$ into a black-and-white topology represented by $\ten{\rho}^\mathrm{phys}$, we employ \textit{projection} filtering, see \cref{eq_topopt14}. To achieve a desired transition to the black-and-white design, \ref{filter} also discusses a gradual increase of the projection filter strength parameter $\psi$ over 10 optimisation loops.

Recalling \cref{eq_topopt06a}, we notice that the cost function is defined in terms of the physical (i.e., projected) densities $\ten{\rho}^\mathrm{phys}$, which depend through the projection filter in \cref{eq_topopt14} on the weight-averaged densities $\ten{\rho}^\mathrm{avg}$, with these depending on the design variables $\ten{\rho}$ through the \mbox{PDE} filter in \cref{eq_topopt13}. Consequently, the following chain rule must be applied while computing the gradients (see \ref{app_sensitivity}):
\begin{equation} \label{eq_topopt15}
    \dfrac{d{\left(\mathrm{cost}\right)}}{d{\ten{\rho}}} = \dfrac{d{\left(\mathrm{cost}\right)}}{d{\ten{\rho}^\mathrm{phys}}} \dfrac{d{\ten{\rho}^\mathrm{phys}}}{d{\ten{\rho}^\mathrm{avg}}} \dfrac{d{\ten{\rho}^\mathrm{avg}}}{d{\ten{\rho}}} \, .
\end{equation}

With the construction above and the definition in \cref{eq_topopt05},
the cost is a non-convex and non-linear function of the design variables $\ten{\rho}$. This implies the existence of multiple local minima for the optimisation problem, highlighting the influence of the initial guess for domain initialisation. We investigate this influence in \ref{app_init}.

\subsection{Robust topology optimisation to ensure a minimum length scale}
\label{robust_topopt}

Weight-averaged density filtering introduces a minimum length scale by means of regularisation. Projection filtering has the opposite effect 
and often leads to the formation of tiny material/void regions \citep{Sigmund2009}. The existence of a minimum length scale, a property referred to as \textit{local} mesh convergence in the jargon of topology optimisation, prevents the formation of impractically small topological features which are susceptible to premature failure upon mechanical tests.

A promising method to ensure local convergence 
is the so-called \textit{robust} formulation of topology optimisation \citep{Sigmund2009,Wang2011}. Based on this method, three sets of projected densities, denoted as eroded $\ten{\rho}_{(-)}^\mathrm{phys}$, intermediate $\ten{\rho}^\mathrm{phys}$, and dilated $\ten{\rho}_{(+)}^\mathrm{phys}$ physical densities, generated by applying projection filtering through \cref{eq_topopt14} with thresholds $\phi_{(-)}$, $\phi$, and $\phi_{(+)}=1-\phi_{(-)}$, are employed in a robust cost function. The difference between the three designs generated by three different thresholds defines manufacturing error bounds on both solid and void phases in the optimised topology. The robust optimisation problem including the contribution of all three sets of projected densities is defined as follows \citep{Wang2011}:
\begin{subequations} \label{eq_topopt16}
\begin{gather}
    \underset{\ten{\rho}}{\mathrm{min}}\!: \, \mathrm{max} \!
    \left( \mathrm{cost}{\left({\ten{\rho}_{(-)}}^\mathrm{phys},\ten{U}_{(-)}\right)}, \,             \mathrm{cost}{\left({\ten{\rho}}^\mathrm{phys},\ten{U}\right)}, \,
           \mathrm{cost}{\left({\ten{\rho}_{(+)}}^\mathrm{phys},\ten{U}_{(+)}\right)}\right) \, , \label{eq_topopt16a} \\
    s.t.: \left\{ \begin{array}{{l}}
        {\ten{K}_{(-)} \ten{U}_{(-)} = \ten{F}} \, , \\
        {\ten{K} \ten{U} = \ten{F}} \, , \\
        {\ten{K}_{(+)} \ten{U}_{(+)} = \ten{F}} \, , \\
        {\dfrac{1}{n_e}\sum_{e=1}^{n_e}{\rho_{e,{(+)}}^\mathrm{phys}} = V_m \dfrac{V_{(+)}}{V}} \, , \\ 
        {0 \leq \ten{\rho} \leq 1} \, ,
        \end{array} \right. \label{eq_topopt16b}
\end{gather}
\end{subequations}
\noindent
where the displacements $\ten{U}_{(-)}$, $\ten{U}$ and $\ten{U}_{(+)}$ are obtained by performing three separate FE analyses with respective global stiffness matrices $\ten{K}_{(-)}$, $\ten{K}$ and $\ten{K}_{(+)}$ for the three sets of physical densities. The volume constraint in \cref{eq_topopt16b} is applied to the dilated physical densities; this is introduced in \cite{Wang2011} to eliminate the numerical artefacts observed in \cite{Sigmund2009} due to using intermediate densities directly. On the right-hand side of the volume constraint, the material volume fraction $V_m$ is multiplied by the ratio of the dilated-to-intermediate volume fractions such that the intermediate densities still follow $V_m$. At the end of the optimisation process, the intermediate physical densities $\ten{\rho}^\mathrm{phys}$ represent the robustly optimised topology. Here, we use $\phi_{(-)} = 0.25$, $\phi = 0.50$ and $\phi_{(+)}=0.75$. Note that the minimum length scale introduced by the robust approach increases as $\abs{\phi - \phi_{(-)}}$ grows. Also, the radius of weight-averaged density filtering directly influences the minimum length scale, see \cite{Wang2011} for details.
\subsection{Optimisation algorithm}
\label{algorithm}
The optimisation algorithm is responsible for updating the design densities in a way that leads to cost minimisation while satisfying the constraints. Having a number of design variables in the order of $10^5$, we choose a gradient-based optimisation strategy and employ MATLAB's general-purpose, gradient-based, constrained optimiser \textit{fmincon} with its \textit{interior-point} method. This method solves a sequence of approximate minimisation problems constructed by adding up the original cost with weighted logarithmic barrier functions, each dependent on a slack variable, with as many slack variables as inequality constraints. The slack variables are adopted to transform the inequality constraints into equality constraints. Meanwhile, the slack variables are restricted to be positive to keep the iterates in the \textit{interior} of the feasible region. As the weighting factor of the barrier functions decreases to zero, the minimum of the approximate problem approaches the minimum of the original cost function. The equality constraints are incorporated with the help of Lagrange multipliers. The approximate problem is then solved using either a direct Newton step or a conjugate gradient step using a trust region method. The default is the former, and if it fails (e.g. due to non-convexity near the current iterate), the algorithm resorts to the latter, see 
\cite{MATLAB_Optimization_Toolbox} for further details.
We provide as inputs 
first-order analytical derivatives (i.e., sensitivities) of the cost with respect to the design variables, which we derive via the adjoint method in \ref{app_sensitivity}. %for a detailed derivation of analytical sensitivities of our cost function via the adjoint method. 
The second-order derivatives of the cost function are approximated by \textit{fmincon} via the \mbox{LBFGS} approach \citep{MATLAB_Optimization_Toolbox}.

To begin the optimisation process, we need an initialisation guess for the design densities $\ten{\rho}$ that satisfies the volume constraint applied on $\ten{\rho}^\mathrm{phys}$. \cref{fig_02_b} shows the initialisation setup with evenly distributed densities, commonly used as a starting guess in topology optimisation and adopted here as the default choice. \Cref{fig_02_c,fig_02_d} show alternative initialisations, which are employed as initial guesses in \ref{app_init}. 
Furthermore, the stopping criterion in control of the termination of the optimisation loop comprises two conditions:
the maximum relative change in the design densities falling below $10^{-4}$, and the maximum number of iterations exceeding $50$.
Whichever of these two conditions is met, the optimisation loop terminates. In our experience, these conditions allow the topology to evolve gradually without many idle iterations.

\begin{figure}[tb]
    \centering
    \includegraphics[width=0.25\linewidth]{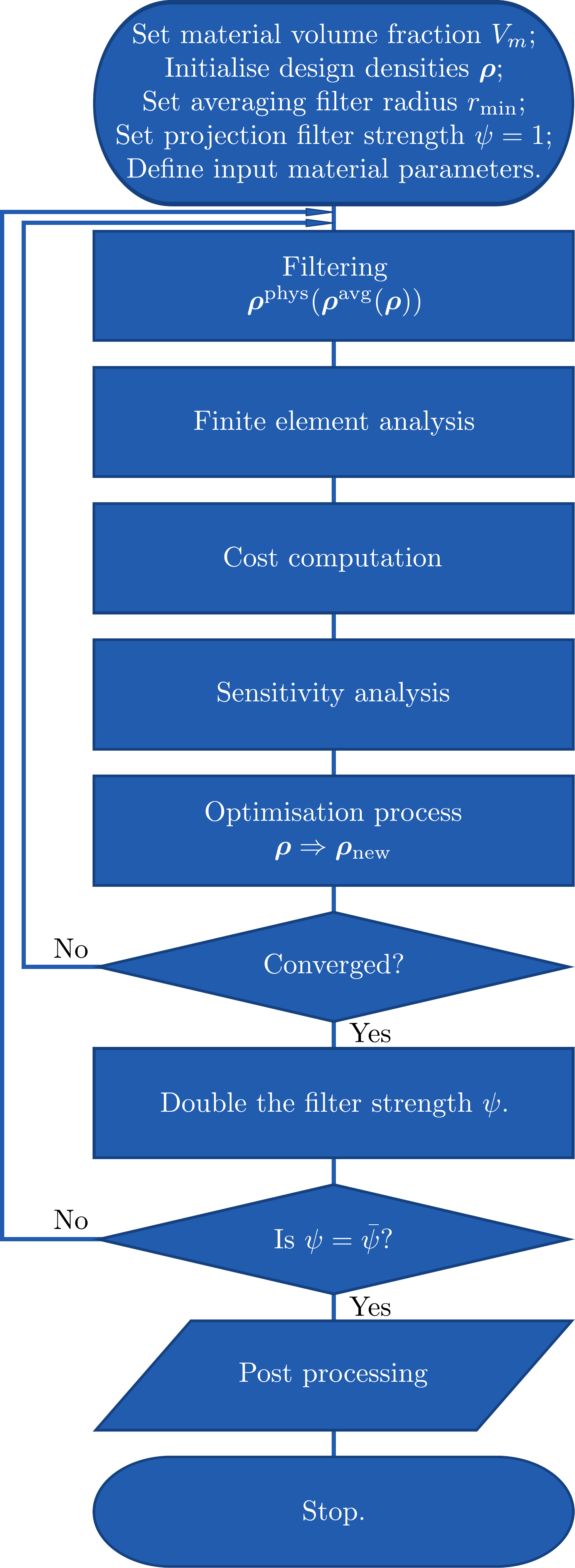}
    \caption[]{Flowchart of the overall optimisation algorithm with gradual increase of the strength of the projection filter (adapted from \cite{Bendsøe2004}).}
    \label{fig_04}
\end{figure}

\cref{fig_04} presents the flowchart of the optimisation algorithm. As per this flowchart, to start the optimisation algorithm, the material volume fraction $V_m$, initial design densities $\ten{\rho}$, the filtering radius $r_\mathrm{min}$, the initial projection filter strength $\psi=1$, and the material parameters $\ten{\theta}$ must be defined. Then, the filtering operation takes place, providing us with the actual topology made by $\ten{\rho}^\mathrm{phys}$. Next, the \mbox{FE} analysis is performed to find the displacement and strain fields which are handed over to the routine which performs the cost and sensitivity computation. The optimiser engine then recruits this information to update the design densities in the direction of the cost minimisation while respecting the constraints. The algorithm keeps iterating until meeting the stopping criterion, thereby doubling the filter parameter $\psi$ to help binarify the topology, and hence continuing until $\psi = \bar{\psi}$ to approach a black-and-white design. Eventually, in case of the presence of minor grey scales in the optimised topology, a post-processing, hard-thresholding step is performed which outputs a sharp $0$-$1$ design.

\section{Results and discussion}
\label{results_discuss}

In this section, we present the main results of our topology optimisation framework to design optimal specimens for noise-tolerant identification of constitutive parameters in isotropic and anisotropic linear elasticity. Further results are reported in the Appendices.
\subsection{Optimisation process and global (mesh) convergence for isotropic elasticity}
\label{convergence}

\begin{figure}[htbp]
    \centering
    \begin{subfigure}{\textwidth}
        \centering
        \includegraphics[width=1\linewidth]{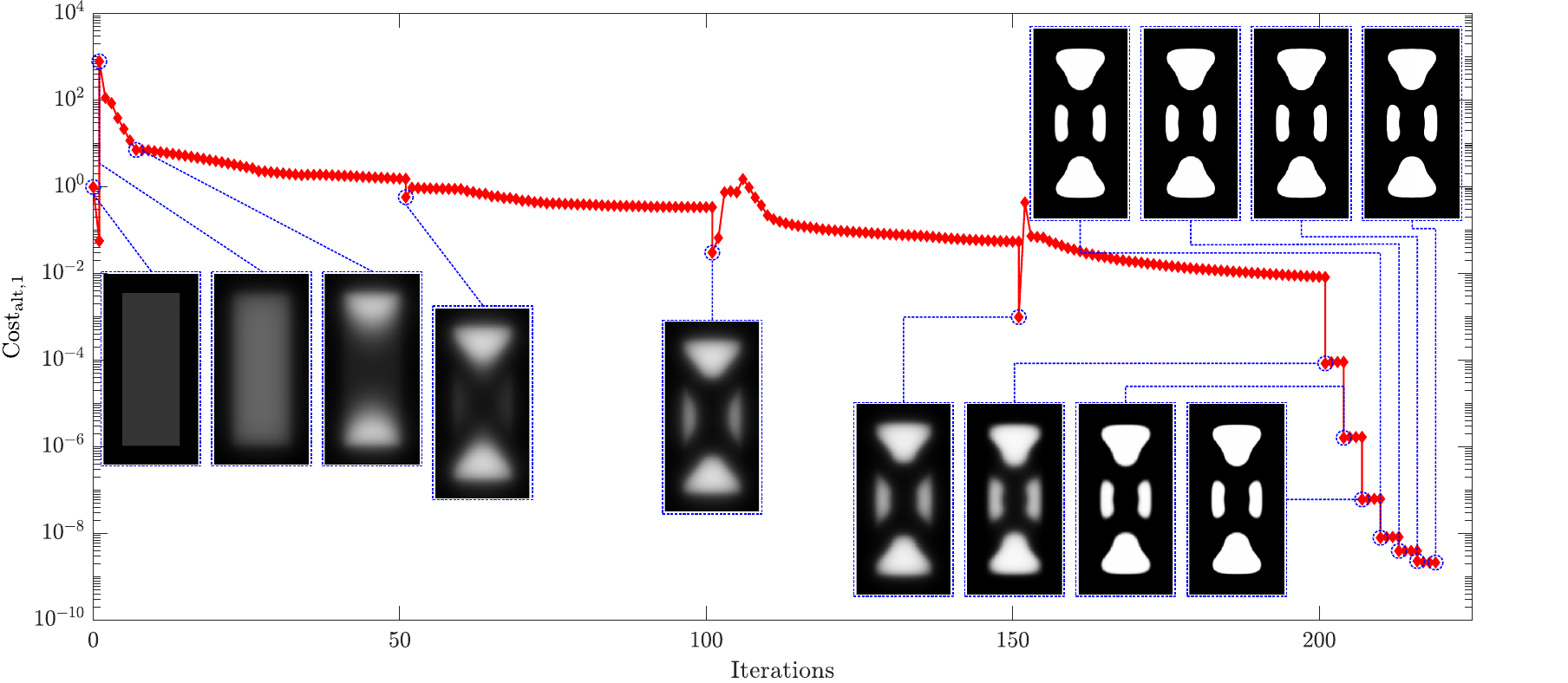} \caption{} \label{fig_05_a}
    \end{subfigure}
    \begin{subfigure}{\textwidth}
        \centering
        \includegraphics[width=1\linewidth]{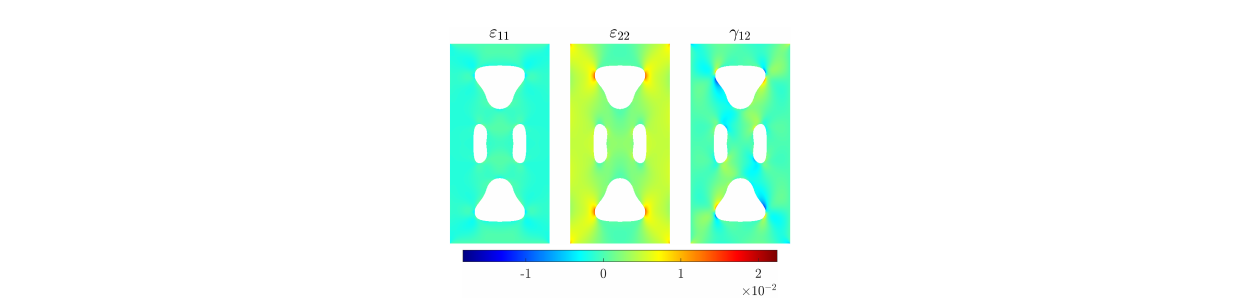} \caption{} \label{fig_05_b}
    \end{subfigure}
    \begin{subfigure}{\textwidth}
        \centering
        \includegraphics[width=1.005\linewidth]{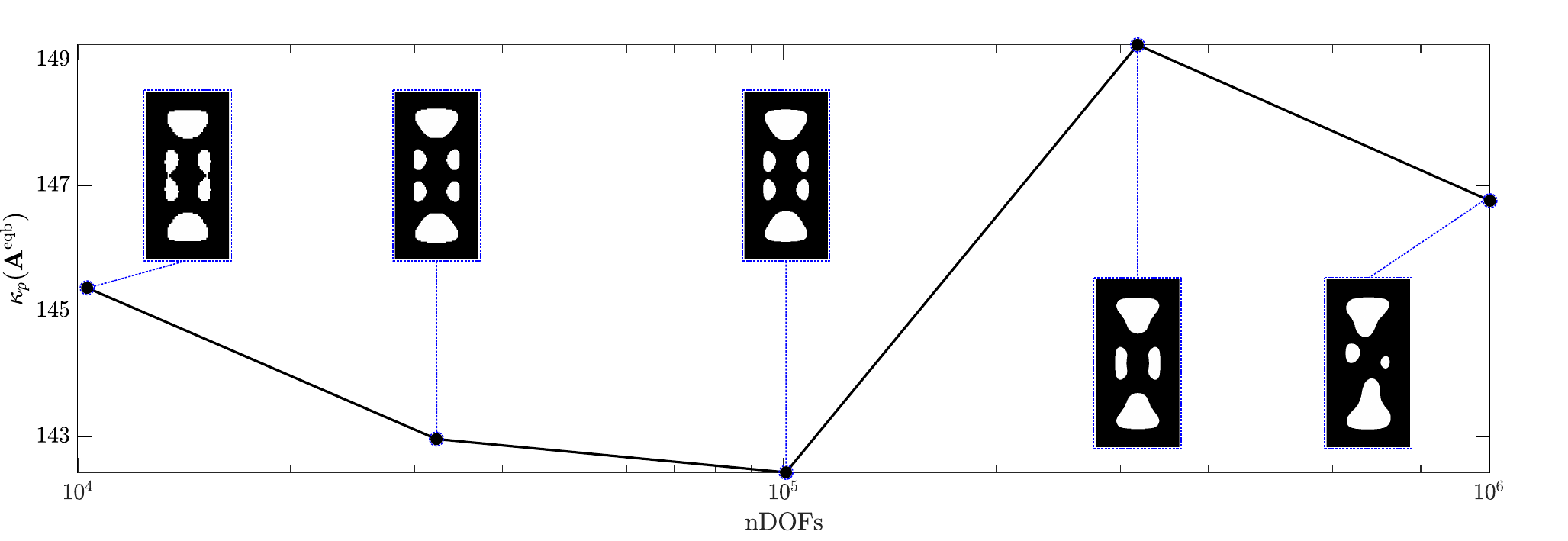} \caption{} \label{fig_05_c}
    \end{subfigure}
    \caption[]{(a) Evolution of the cost function in \cref{eq_topopt04} and of the topology during optimisation; (b) contours of the strain components for the optimised topology; (c) final cost and optimised topology versus mesh refinement.}
    \label{fig_05}
\end{figure}

\begin{figure}[htbp]
    \centering
    \begin{subfigure}{\textwidth}
        \centering
        \includegraphics[width=1\linewidth]{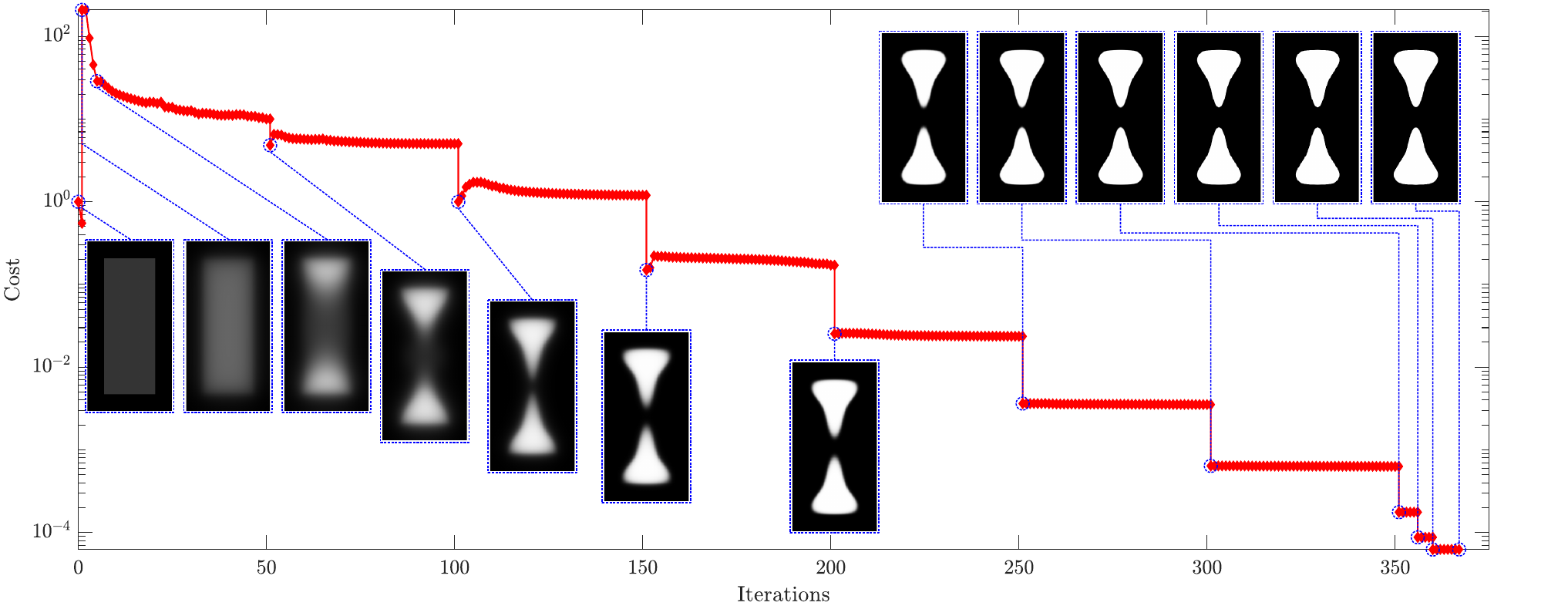} \caption{} \label{fig_06_a}
    \end{subfigure}
    \begin{subfigure}{\textwidth}
        \centering
        \includegraphics[width=1\linewidth]{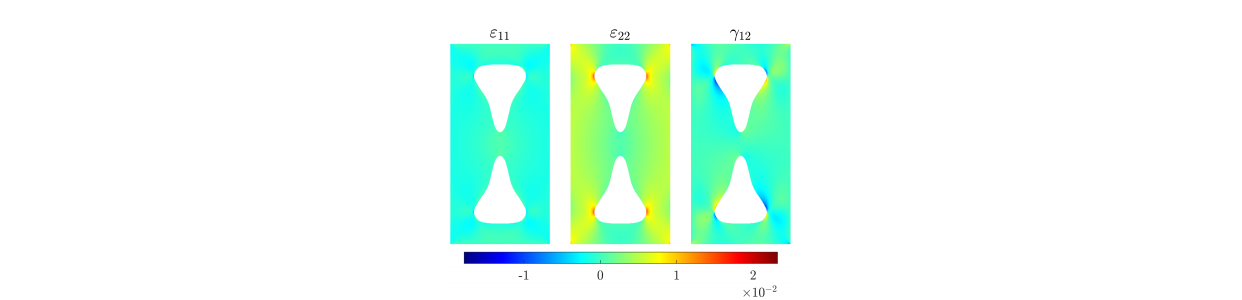} \caption{} \label{fig_06_b}
    \end{subfigure}
    \begin{subfigure}{\textwidth}
        \centering
        \includegraphics[width=1.005\linewidth]{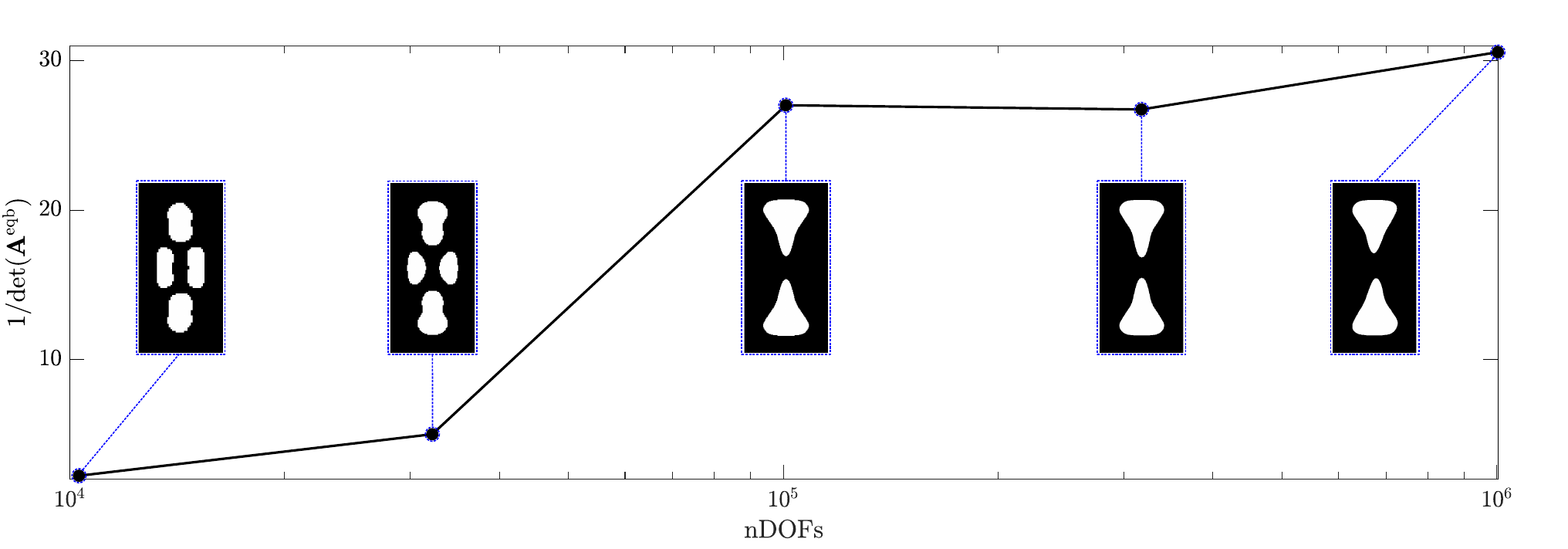} \caption{} \label{fig_06_c}
    \end{subfigure}
    \caption[]{(a) Evolution of the cost function in \cref{eq_topopt05}) and of the topology during optimisation; (b) contours of the strain components for the optimised topology; (c) final cost and optimised topology versus mesh refinement.}
    \label{fig_06}
\end{figure}

This section examines the performance of the cost functions in \cref{eq_topopt04} (with $p=8$) and \cref{eq_topopt05}
by analysing their optimisation process, the strain fields in their output topology subjected to the tensile test of Fig. \ref{fig_02}, and their mesh convergence behaviour. The results from the cost functions in \cref{eq_topopt04} and \cref{eq_topopt05} are given in \cref{fig_05} and \cref{fig_06}, respectively. An isotropic material with Young's modulus $E=200~\mathrm{GPa}$ and Poisson's ratio $\nu=0.3$ is assumed for this analysis.

\cref{fig_05_a,fig_06_a} illustrate the topology optimisation process. 
Each plot shows the evolution of the cost (red diamonds) and topology against the number of iterations during optimisation. 
Each optimisation process is initialised with evenly distributed densities (as shown earlier in \cref{fig_02_b}), and a cost of one due to the normalised definition. 
In the next iteration, the weight-averaging \mbox{PDE} filter becomes active, which induces some blur in the topologies and a sharp jump in the cost value. 
Shortly after, there is a rapid drop in the cost until iteration $\#7$ in \cref{fig_05_a} and iteration $\#4$ in \cref{fig_06_a}, where an early-stage picture of the topology emerges. 
At iteration $\#51$, the algorithm reaches the maximum number of iterations per optimisation loop
and 
the projection-filter parameter $\psi$ is doubled, which stimulates stronger 0-1 binarification as well as a sharp drop in the cost value. The algorithm continues similarly with the next optimisation loops, gradually leading to a 
black-and-white topology. 
Overall, both cost functions 
exhibit mostly decreasing values during the course of optimisation (except for a few iterations). After four loops in \cref{fig_05_a} and seven loops in \cref{fig_06_a}, each consisting of $50$ iterations, the changes in the topologies become rather negligible, thereby activating the other stopping criterion, i.e., the lower bound on the maximum relative change in the design densities, resulting in a more rapid termination of the remaining loops. The final optimised topologies are delivered at the end of loop $\#10$ (when $\psi = \bar{\psi}$). To speed up the optimisation process, a possibility could be setting the stopping criteria adaptively to allow less strict termination of the algorithm in the last 4-5 optimisation loops, as they are mainly responsible for increasing the contrast of the design, i.e., binarification. However, to ensure a smooth and gradual topology update, we have not incorporated such adaptive stopping criteria. 
To quantify the discreteness of the optimised topologies, \cite{Sigmund2007} define a grey-level index as
\begin{equation} \label{eq_results01}
    g_\mathrm{idx} = \dfrac{4}{n_e} \sum_{e=1}^{n_e} {\rho_e^\mathrm{phys} \left(1-\rho_e^\mathrm{phys} \right)} \, .
\end{equation}
\noindent
Designs with $g_\mathrm{idx} < 1\%$ are recognised as sufficiently discrete. Using this metric, our optimised topologies in \cref{fig_05_a} and \cref{fig_06_a} yield $g_\mathrm{idx}= 0.39\%$ and $0.20\%$, respectively. The optimised designs finally undergo post-processing with hard thresholding to ensure $g_\mathrm{idx}=0$ in the final output. The comparison among the two cost functions shows a steadier behaviour (i.e., declining more uniformly through longer-lasting optimisation loops) as well as a smaller final grey-level index for the cost function in \cref{eq_topopt05}. 
Both optimised topologies contain two symmetric holes in the top and bottom halves of the specimen, and the one reached by the cost function in \cref{eq_topopt04} contains two more holes. The $2$-norm condition number is similar for both designs; it is given by $\kappa_2 = 193$ and $204$ for the topologies in \cref{fig_05_a} and \cref{fig_06_a}, respectively.

\cref{fig_05_b,fig_06_b} plot the contours of the strain components $\varepsilon_{11}$, $\varepsilon_{22}$ and $\gamma_{12}$. 
For the loading conditions in \cref{fig_02_b}, $\varepsilon_{11}$ is almost zero in both specimens, whereas $\varepsilon_{22}$ shows concentrations and gradients at the side corners of the holes and at the four outer corners. Similarly, the shear strain also exhibits some concentrations. The ranges of strain values for the two specimens are almost identical. However, the two additional side holes in \cref{fig_05_b} generate some minor gradients of the shear strain component, a favourable feature missing in \cref{fig_06_b}. 
In the next section we provide quantitative measures to highlight the importance of the strain patterns generated as a result of topology optimisation.

\cref{fig_05_c,fig_06_c} investigate the \textit{global} convergence of the optimised topologies as the total number of DOFs (nDOFs) increases. 
The determinant-based cost function leads to acceptable mesh convergence, with only minor changes as the FE grid is refined. 
The slight distortion of the holes at $\mathrm{nDOFs} \approx 10^{6}$ is likely due to the effect of roundoff error. 
The mesh refinement level with $\mathrm{nDOFs} \approx 10^{5.5} \approx 3.2 \times 10^5$ (yielding $n_e \approx 1.6 \times 10^5$ design variables) is considered sufficient for convergence, and all further analyses are based on it.
Conversely, $\kappa_p{(\ten{A}^\mathrm{eqb})}$ renders an inferior convergence behaviour since the optimised topologies continue to change while the FE mesh is refined. 
 We note that the consistent definition of the weighting parameters $\lambda_r$ and $\lambda_q$ (as discussed in \ref{app_weighting_factor}) is of crucial importance to achieve global convergence in the optimised topologies. 
Considering all the aspects discussed above, the determinant-based cost function from \cref{eq_topopt05} proves to be the best choice.
This statement continues to hold also for anisotropic materials, as exemplified later. %in \cref{fig_07,fig_08} versus \cref{fig_app02}. 
Therefore, from now on we only focus on the determinant-based cost function in \cref{eq_topopt05}. 
In \ref{app_other_cost} we provide more results for the cost function in \cref{eq_topopt04} and explore a few additional cost definitions.

\subsection{Topology optimisation for material parameter identification in orthotropic elasticity}
\label{orthotropic}

As mentioned in \cref{problem,algorithm}, the topology optimisation framework requires the material parameters as inputs to be able to solve for the nodal displacements through the \mbox{FE} analysis in \cref{eq_topopt06b}. This may sound problematic, as the material parameters are the unknowns of the parameter identification for which the specimen is to be designed. In this section, we aim to investigate the effect of the input material parameters on the output topologies.
To this end, we focus our attention on orthotropic materials under the plane stress assumption. Such materials exhibit different stiffness properties along two orthogonal in-plane directions (material local coordinates), as shown in \cref{fig_01}. We denote with $x$ the strong material direction, %or the anisotropy orientation,
situated at an angle $\beta$ %measured counterclockwise 
from the global coordinate $1$, and with $y$ the weak material direction. 
To characterise the planar behaviour of orthotropic materials, four parameters are required: the longitudinal Young's modulus, $E_{xx}$, the transverse Young's modulus $E_{yy}$, the in-plane shear modulus $G_{xy}$, and the in-plane Poisson's ratio $\nu_{xy}$ (or the transverse Poisson's ratio $\nu_{yx} = \nu_{xy} \, E_{yy} / E_{xx}$). To characterise the intensity of anisotropy, we utilise the two dimensionless parameters introduced in \cite{Nejati2019}: 
\begin{equation} \label{eq_results02}
    \alpha_1 = \dfrac{E_{xx}}{E_{yy}} \, \, \, \, \, \, \, \,
    \alpha_2 = \dfrac{G_{xy}}{G_{xy}^{sv}} \, ,
\end{equation}
with
\begin{equation}
    \dfrac{1}{G_{xy}^{sv}} = \dfrac{1}{E_{xx}} + \dfrac{1}{E_{yy}} + \dfrac{2\nu_{xy}}{E_{xx}} \, . 
\end{equation}
Note that for an isotropic material $\alpha_1=\alpha_2=1$. 
Based on the analysis in \cite{Nejati2019}, the constitutive response is predominantly governed by the anisotropy ratios $\alpha_1$, $\alpha_2$ and the anisotropy angle $\beta$, whereas the influence of $\nu_{xy}$ is negligible. Thus, we consider the range of material parameters in \cref{tab_results01}, representing a broad spectrum of orthotropic materials, to be given as inputs to the topology optimisation framework. 
Due to the normalisation of the cost function, we expect only these dimensionless ratios to affect the results.
\begin{table}[H]
\caption{Material properties to define a range of orthotropic materials.} 
\label{tab_results01}
\centering
\renewcommand\arraystretch{1.1}
\begin{tabular}{l@{\hspace{20pt}}l@{\hspace{20pt}}l@{\hspace{20pt}}l}
\hline
    $\alpha_1$ & $\alpha_2$ & $\beta$ ($\degree)$  & $\nu_{xy}$ \\
\hline
    $\{4,8,\ldots,20\}$ & $\{0.5,0.75,\ldots,1.5\}$ & $\{0,15,\ldots,90\}$ & $0.3$  \\
\hline
\end{tabular}  
\end{table}

\begin{figure}[tb]
    \centering
    \includegraphics[width=1\linewidth]{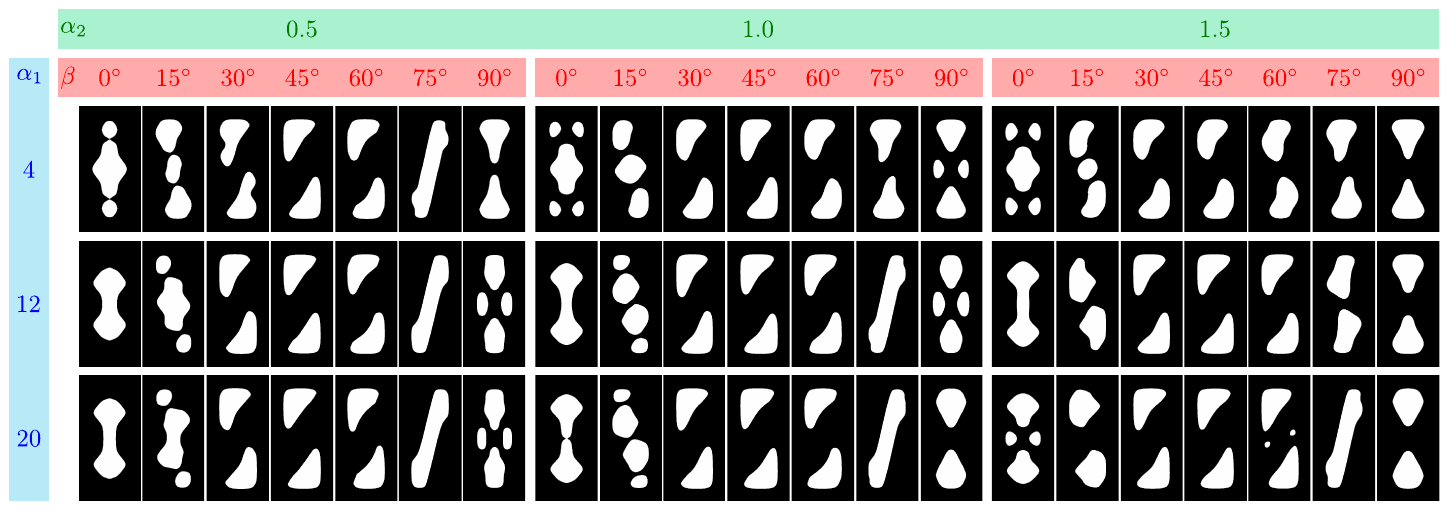}
    \caption[]{Optimised topologies for different orthotropic material parameters in input. The densities are initialised evenly, and the material volume fraction is set to $V_m = 80\%$.}
    \label{fig_07}
\end{figure}

\cref{fig_07} portrays the topologies generated using in input some of the orthotropic material parameters from \cref{tab_results01} and a material volume fraction of $V_m=80\%$. 
The first observation is that the choice of the input material affects the optimised topology, with %Looking at the specimen topologies more closely, it is realised that 
the most significant changes 
resulting from the variation of the anisotropy angle $\beta$. 
We later use this information to quantify the performance of selected topologies.
Comparing \cref{fig_07} with \cref{fig_06} for the isotropic case, we note that similar holes form when the strong material direction coincides with the loading axis ($\beta=90\degree$). As the anisotropy angle decreases towards $0\degree$, the holes tend to deform accordingly. Along with this deformation, the holes may merge or decompose into smaller pieces. These changes result from a complex interplay between the anisotropy orientation and the ensuing strain field (which determines $\det{\!(\ten{A}^\mathrm{eqb})}$), the weight-average filtering, the presence of the outer passive frame, and the constraint of fixed material volume fraction. Another prominent feature in the optimised topologies is their (bisect-dual-flip) symmetry,
which is a consequence of symmetry in $\ten{A}^\mathrm{eqb}$, symmetric loading configuration, and evenly distributed initialisation densities. It is shown in \ref{app_init} that initialisations with sufficient randomness can disrupt such symmetric patterns. Lastly, we observe that a few topologies
contain narrow strips of material connecting the holes, which may prematurely fail upon loading the specimen in a real experiment. In \cref{results_robust_topopt} we discuss the application of robust topology optimisation as a remedy to this issue.

\begin{figure}[htb]
    \centering
    \includegraphics[width=1\linewidth]{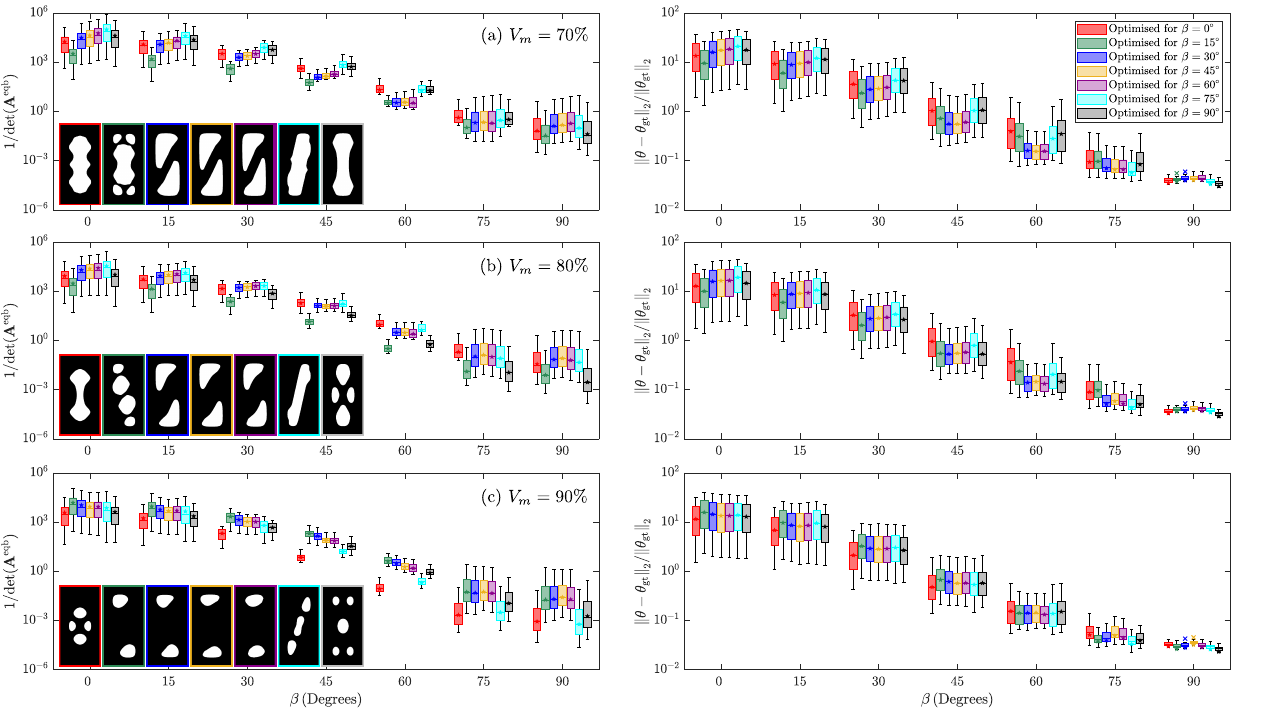}
    \caption[]{Identification of the orthotropic material parameters using optimised topologies obtained with $(\alpha_1,\alpha_2)=(12,1.0)$ with different volume fractions. The left plots visualise the unnormalised cost (in $mm^{-12}$) while the right ones report the identification error (in percentage), both against the anisotropy angle $\beta$. The boxes and whiskers represent the range of results for different anisotropy ratios $\alpha_1$ and $\alpha_2$. The cross signs $(\times)$ denote the outliers (i.e. cases lying $1.5\times(\text{interquartile range})$ away from the top or bottom of the boxes) and the star signs $(\star)$ designate the cases which have the same $\alpha_1$ and $\alpha_2$ value as the optimised topologies.}
    \label{fig_08}
\end{figure}

We now assess the performance of the optimised specimen topologies 
for identification of the material parameters $\ten{\theta}$. Note that $\ten{\theta}$ contains $n_f=6$ unknown stiffness components as in \cref{eq_cost06b}, since the material is assumed to be linearly elastic with no information on the type of material symmetry. 
We use artificially generated deformation fields, perturb the strains at the integration points with Gaussian noise with zero mean and standard deviation $\gamma_f = 10^{-3} \times \bar{u}/L_Y$, where $\bar{u}/L_Y$ is the nominal strain applied to the specimen, and perform no denoising. We apply the already created topologies to identify the elastic material properties which can be generated via the inputs in \cref{tab_results01}. 
We select as representative the seven topologies generated by $(\alpha_1,\alpha_2)=(12,1.0)$ from \cref{fig_07}, and investigate their effectiveness for parameter identification. Further, to provide some insight on the effect of the material volume fraction, we generate the respective set of topologies with $V_m = 70\%$ and $V_m = 90\%$. \cref{fig_08} 
illustrates the results, in terms of
 cost (without normalisation) and
 parameter identification error ${\norm{\ten{\theta}-\ten{\theta}_\mathrm{gt}}}_2/{\norm{\ten{\theta}_\mathrm{gt}}}_2$, where 
$\ten{\theta}_\mathrm{gt}$ is the vector of the ground-truth material parameters. %whereas $\ten{\theta}$ is the gathering of identified material parameters obtained through \cref{eq_cost17} using the noisy strain field. 

The most prominent trend observed in \cref{fig_08} is the overall decrease of both the cost and the identification error when the anisotropy angle varies from $\beta = 0\degree$ to $90\degree$ under uniaxial loading, i.e. as 
the strong material direction 
comes closer to the loading direction, leading to a better identification of 
$E_{xx}$ 
(since $E_{xx}$ is the largest stiffness component, its identification error affects the $\mathcal{l}_2$-norm error the most significantly). Furthermore, investigation of the individual identification error in each of the stiffness components reveals that, normally, the hardest parameter to be identified for $\beta<45\degree$ is $E_{xx}$, for $\beta>45\degree$ is $E_{yy}$, and for $\beta=45\degree$ is $G_{xy}$. 
Considering the middle-row results pertaining to $V_m = 80\%$ in \cref{fig_08}, we see that the different optimised topologies yield different costs, with the topologies optimised for $\beta = 15\degree$ and $90\degree$ outperforming the rest in almost all cases. As an example, to calibrate the material parameters for orthotropic materials with $\beta = 0\degree$, considering the medians in the boxes and whiskers as representative, the unnormalised cost for the $15\degree$ topology is $2.14\times10^{3}~mm^{-12}$, while for the $0\degree$ topology it is $6.52\times10^{3}~mm^{-12}$. As for the identification error, the $15\degree$ topology delivers the lowest error of $10\%$, while the $0\degree$ topology gives $13\%$. The better performance of the topologies optimised for $\beta = 15\degree$ and $90\degree$ can be explained with the closely spaced holes appearing in these two specimens, which cause localised strain gradients adjacent to the holes.
Considering the results for $V_m = 70\%$ and $90\%$, a similar superiority in performance is observed for the topologies created with $\beta = 15\degree$ in the former, and with $\beta = 0$, $15$ and $90\degree$ in the latter, again due to the effect of tightly arranged holes. Overall, it is unexpected that certain topologies be superior to others in almost all cases. We would rather expect that every topology produced for a material with a specific $\beta$ be the optimal topology for identifying materials with that value of $\beta$. A possible justification is the non-convexity of the cost function and the effect of initialisation which can lead the optimisation to a local minimum, as illustrated in more detail in \ref{app_init}. A further reason may be the projection filtering 
which induces abrupt changes in the optimised topologies as the input anisotropy angle varies between $0\degree$ and $90\degree$.
We shed more light on this in \cref{results_robust_topopt}.
Comparing the results for different material volume fractions in \cref{fig_08}, it is evident that the optimised topologies remain quite consistent as $V_m$ changes. With the increase of $V_m$, the holes may shrink and decompose into smaller holes, or coalesce and form fewer but larger holes. The lowest identification error bounds belong to $V_m = 90\%$. A possible explanation is that a higher material volume fraction enriches the system of equations in \cref{eq_cost17} with more material points. %which in turn leads to a more robust solution. 
However, the limit case of $V_m = 100\%$ (not shown here) expectedly leads to exploding identification errors. The plot for $V_m = 90\%$ shows that the performance differences between the optimised topologies are rather small, hence the topologies optimised with different anisotropy angles perform (almost) equally well in the material identification process. 
This is an important point as it diminishes the significance of defining accurate material parameters as inputs to the topology optimisation framework. Here, by the definition of input material parameters, we simply mean defining the anisotropy ratios $\alpha_1$, $\alpha_2$ and the anisotropy orientation $\beta$, while the actual individual stiffness components gathered in $\ten{\theta}$ are not required. Using the actual material parameters to design the test specimen contradicts the purpose of the constitutive law calibration, yet it has not been addressed in the literature \citep{Chamoin2020,Almeida2020,Barroqueiro2020,Goncalves2023}.  
Based on the above practically admissible results, we would recommend running the topology optimisation algorithm for a few anisotropy orientations (e.g. $\beta = 0, 15$ and $90\degree$) with arbitrary $\alpha_1$ and $\alpha_2$ values, and then pick the specimen design with the largest number of holes, which is expected to enable robust identifiability for any 2D orthotropic material.

\begin{figure}[htb]
    \centering
    \includegraphics[width=0.6\linewidth]{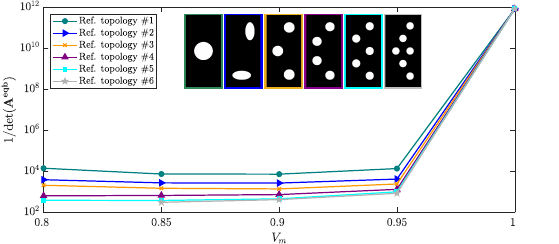}
    \caption[]{The dependence of the unnormalised cost (in $mm^{-12}$) on the material volume fraction investigated for six reference topologies.}
    \label{fig_12}
\end{figure}

We provide a more explicit representation of the dependence of the cost function on the material volume fraction in \cref{fig_12}. Here, we plot $1/\det{\!(\ten{A}^\mathrm{eqb})}$ versus the material volume fraction in the range $V_m = 80-100\%$ for six simple topologies created manually and numbered according to the number of holes. The reason behind choosing simple topologies is the ability to reproduce them exactly (to exclude the influence of shape change) with different volume fractions. It is not feasible to account for low volume fractions as they lead to invalid topologies, sometimes causing shape changes with merging holes, or introducing narrow strips of material (as also in the case of topology \#6 at $V_m = 80\%$), which are practically avoided in topology optimisation through filtering. The material under study is orthotropic with $(\alpha_1,\alpha_2)=(12,1.0)$, and the plots in \cref{fig_12} reflect the average results for different anisotropy angles $\beta \in \{0,15,\ldots,90\}$. Material volume fractions of $ 
85\% \leq V_m \leq 90\%$ are observed to be optimal irrespective of the topology used, probably because such volume fractions lead to sufficient richness of the system of equations due to many material points while preserving a heterogeneous design with non-uniform strain maps. Increasing the volume fraction further has a negative impact, whereby the extreme case of the full plate (i.e., $V_m = 100\%$) has a dramatically higher cost due to its uniform strain distribution, hence loss of identifiability. On the other hand, the reduction of the volume fraction depletes the information available on the deformation field, yielding less accurate material parameters.

\begin{figure}[H]
    \centering
    \begin{subfigure}{\textwidth}
        \centering
        \includegraphics[width=1\linewidth]{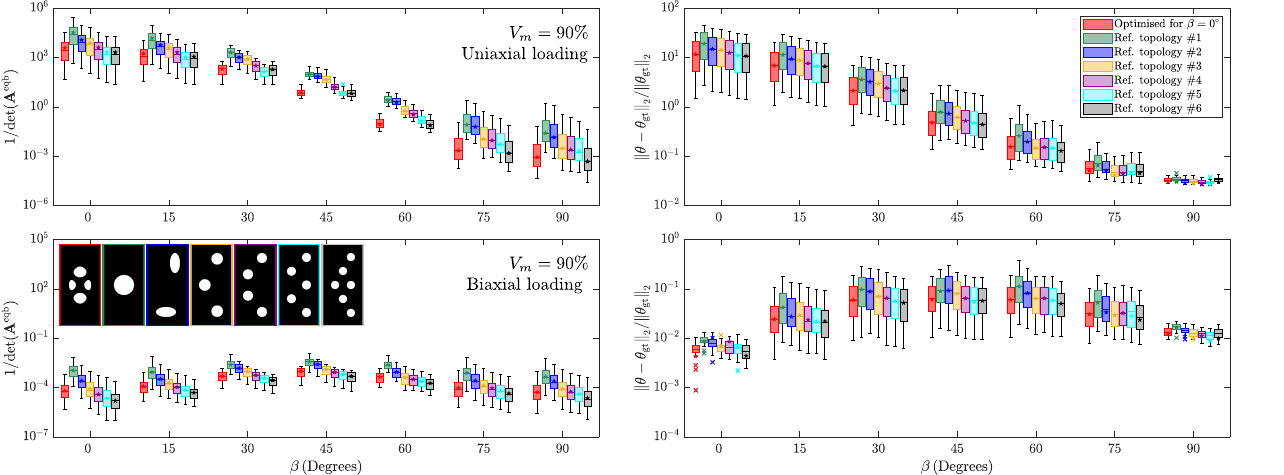} \caption{} \label{fig_09_a}
    \end{subfigure}
    \begin{subfigure}{\textwidth}
        \centering
        \includegraphics[width=1\linewidth]{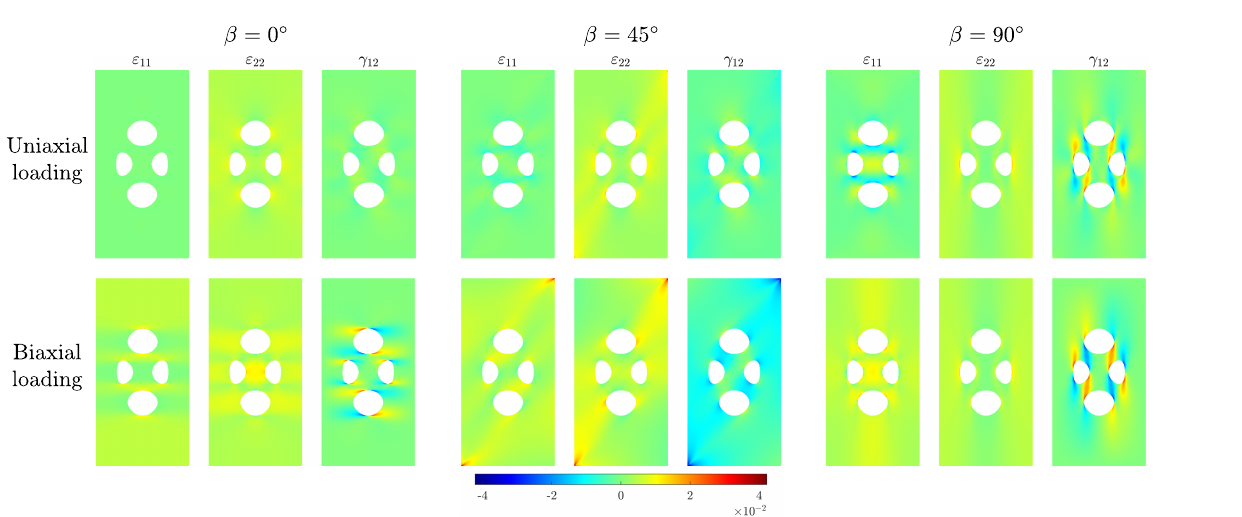} \caption{} \label{fig_09_b}
    \end{subfigure}
    \caption[]{Performance assessment of a chosen topology (produced with $(\alpha_1,\alpha_2,\beta) = (12,1.0,0\degree)$ and $V_m = 90\%$ from \cref{fig_08}) under uniaxial and biaxial tensile loading conditions: (a) comparison of the unnormalised cost (in $mm^{-12}$) and the identification error (in percentage) versus six reference topologies; (b) contours of strain components obtained for anisotropy orientations $\beta = 0, 45$ and $90\degree$.}
    \label{fig_09}
\end{figure}

\cref{fig_09_a} compares the performance between the optimised topology generated with $(\alpha_1,\alpha_2,\beta) = (12,1.0,0\degree)$ and the six reference topologies. All the topologies share the same material volume fraction $V_m = 90\%$. 
\mbox{EUCLID} has previously employed designs similar to reference topologies $\#1$ and $\#2$ to discover different classes of constitutive laws \citep{Flaschel2021a, Flaschel2022,Flaschel2022a}, although none of them concerned orthotropic materials. The remaining reference topologies should trigger more heterogeneous strain patterns as the number of their holes increases. The comparison is made for both uniaxial and biaxial loading (the latter applies an additional horizontal extension $u=\bar{u} \times L_X/L_Y$ to the right edge of the specimen). As shown in \cref{fig_09_a} and expected, the cost and the identification error decrease as the number of holes increases. 
However, results for topologies $\#5$ and $\#6$ are very similar, indicating that a saturation is reached%the consecutive addition of holes does not invariably bring added efficiency, but rather leads to a saturation 
, as also pointed out in \cite{Ihuaenyi2024}.
The selected optimised topology, 
which contains 4 but optimally designed holes, appears as effective as 
the most complex 6-hole manually designed topology, %(whose designs are rather not trivial and require insight into good practice for material identification). 
which is the advantage of the automated design via topology optimisation. 
Overall, the cost and identification error drop by orders of magnitude due to richer strain patterns in the biaxial loading setup (see \cref{fig_09_b}).

\cref{fig_09_b} illustrates the contours of the in-plane strain components for the automatically optimised topology, 
obtained by uniaxial and biaxial numerical experiments with $\beta = 0, 45$ and $90\degree$. 
Under uniaxial loading, the case $\beta = 0\degree$ results in rather uniform strain fields with negligible horizontal and shear strains. For $\beta = 45\degree$, the strain contours acquire directional variations and larger values, and gradients of the shear strain appear. %, whereas the normal strain $\varepsilon_{11}$ appears to remain quite idle. Finally, checking the contours of
For $\beta = 90\degree$, the largest values of the strains and their gradients are obtained. %it is depicted that the variability in the strain field and the emergence of strain gradients 
Changing the loading condition to biaxial further improves the results by generating more heterogeneous strain patterns also for $\beta < 45\degree$. This may explain the discussed trends of cost and identification error observed in \cref{fig_08,fig_09_a} when $\beta$ changes from $0\degree$ to $90\degree$.

\begin{figure}[tb]
    \centering
    \includegraphics[width=1\linewidth]{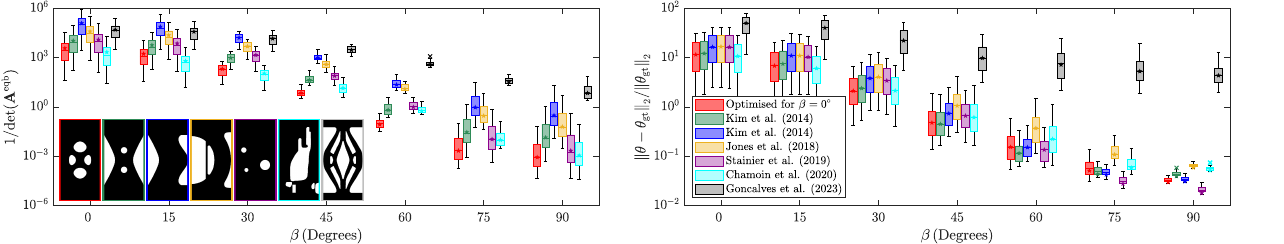}
    \caption[]{Performance assessment of a chosen topology (produced with $(\alpha_1,\alpha_2,\beta) = (12,1.0,0\degree)$ and $V_m = 90\%$ from \cref{fig_08}) versus selected designs from the literature under uniaxial tensile loading. The left plot visualises the unnormalised cost (in $mm^{-12}$) while the right one reports the identification error (in percentage), both against the anisotropy angle $\beta$.}
    \label{fig_13}
\end{figure}

Upon a request by the reviewers, in \cref{fig_13} we compare the performance of our optimised topology (generated with $(\alpha_1,\alpha_2,\beta) = (12,1.0,0\degree)$) against selected topologies designed in the literature \citep{Kim2014,Jones2018b,Stainier2019,Chamoin2020,Goncalves2023}. Such a comparison has to be considered with caution because of two main reasons: (1) each specimen from the literature has been designed for the identification of different and specific material models, from linear elasticity \citep{Stainier2019,Chamoin2020}, to elasto-plasticity \citep{Kim2014,Goncalves2023} and visco-plasticity \citep{Jones2018b}, using different identification techniques; (2) each topology has a specific material volume fraction, size, aspect ratio, resolution, and loading configuration. While bearing in mind the limitations in (1), we have tried our best to address most of the discrepancies in (2) by redesigning the selected topologies with minimal changes (not expected to have considerable influences). This means that the redesigned specimens feature equal size, aspect ratio and resolution as well as loading configuration (i.e., uniaxial tensile loading with full-width grip at both ends), but have different material volume fractions. The comparison between the specimen designs is based on the identification of the linear elastic constitutive parameters produced via the inputs in \cref{tab_results01} through the equilibrium gap method  outlined in \cref{parameter_identification}. The results in \cref{fig_13} reveal that our optimised topology performs very well in comparison to the literature designs, leading to lower costs and identification errors in a wide spectrum of the considered orthotropic materials. The notched 2-hole specimen by \cite{Kim2014} exhibits quite similar performance to our design, while the 3-hole specimen by \cite{Stainier2019} and the optimised specimen by \cite{Chamoin2020} are optimal in the very high and very low anisotropy orientation regimes, respectively. In contrast, the optimised specimen by \cite{Goncalves2023}, probably due to its low volume fraction (as analysed in \cref{fig_08,fig_12}), leads to high identification errors.

\subsection{Robust topology optimisation}
\label{results_robust_topopt}

 In a few cases, the optimised topologies comprise narrow strips of material which are prone to premature failure upon testing. This issue mostly arises due to the application of the projection filter with a single threshold $\phi$ (to achieve a black-and-white design) since this destroys the minimum length scale introduced by weight-averaged density filtering \citep{Wang2011}. To solve this problem, we resort to the robust formulation to topology optimisation outlined in \cref{robust_topopt} to obtain a minimum length scale on the physical (i.e., projected) densities. %In this regard, the optimisation problem expressed in \cref{eq_topopt16} (with its cost including the contributions of three projection levels instead of one) is considered to rule the optimisation process.

\begin{figure}[thb]
    \centering
    \begin{subfigure}{0.45\textwidth}
        \centering
        \includegraphics[width=1\linewidth]{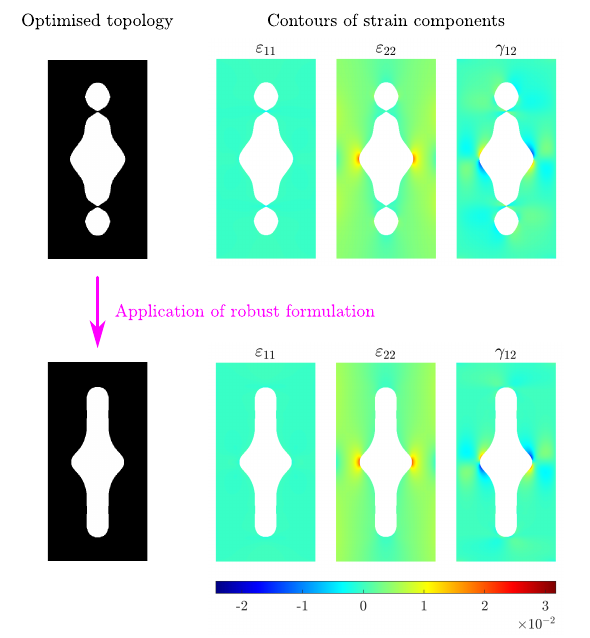} \caption{} \label{fig_10_a}
    \end{subfigure}
    \hspace{5mm}
    \begin{subfigure}{0.45\textwidth}
        \centering
        \includegraphics[width=1\linewidth]{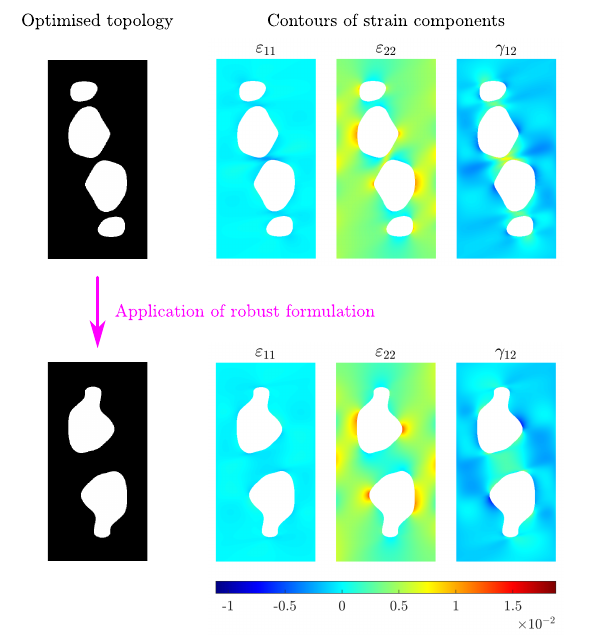} \caption{} \label{fig_10_b}
    \end{subfigure}
    \caption[]{Effect of the \textit{robust} formulation in topology optimisation to avoid tiny topological features: (a) topologies generated with inputs $(\alpha_1,\alpha_2,\beta) = (4,0.5,0\degree)$ ; (b) topologies generated with inputs $(\alpha_1,\alpha_2,\beta) = (20,1.0,15\degree)$.}
    \label{fig_10}
\end{figure}

\cref{fig_10} shows two sample cases from \cref{fig_07}, produced with $(\alpha_1,\alpha_2,\beta) = (4,0.5,0\degree)$ and $(20,1.0,15\degree)$. The original topologies and their strain contours are shown in the upper row, where we note the existence of narrow strips of material and the consequent strain localisation which may lead to premature failure upon loading. Moreover, 
the \mbox{DIC} system fails to capture the deformation field in regions situated in the vicinity of the edges, resulting in the loss of valuable data. As shown in the lower row in \cref{fig_10}, robust topology optimisation can fix the problem, leading to 
merged holes and wider material bands.
\begin{figure}[tb]
    \centering
    \begin{subfigure}{\textwidth}
        \centering
        \includegraphics[width=1\linewidth]{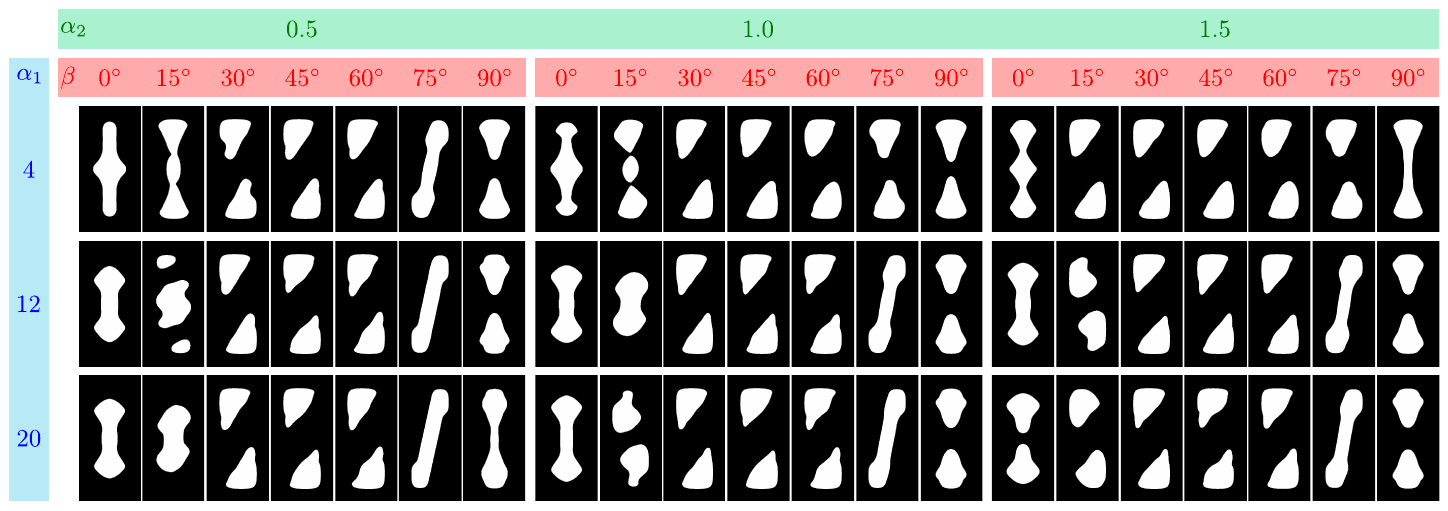} \caption{} \label{fig_11_a}
    \end{subfigure}
    \begin{subfigure}{\textwidth}
        \centering
        \includegraphics[width=1\linewidth]{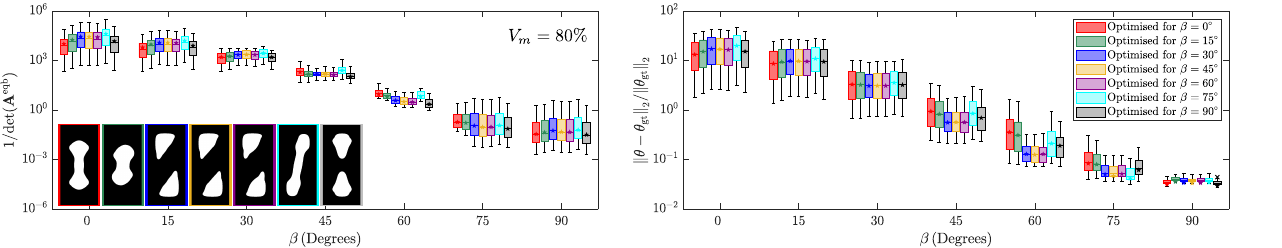} \caption{} \label{fig_11_b}
    \end{subfigure}
    \caption[]{The application of the \textit{robust} formulation in topology optimisation: (a) optimised topologies for different orthotropic materials. The densities are initialised evenly, and the material volume fraction is set to $V_m = 80\%$; (b) identification of the orthotropic material parameters using the optimised topologies obtained with $(\alpha_1,\alpha_2)=(12,1.0)$ from part (a). The left plot visualises the unnormalised cost (in $mm^{-12}$) while the right one reports the identification error (in percentage), both against the anisotropy angle $\beta$.}
    \label{fig_11}
\end{figure}

\cref{fig_11} presents the robustly optimised topologies and the performance of a few selected items in terms of cost and material identification error. 
Comparing the topologies in \cref{fig_11_a} with the 
ones in \cref{fig_07}, it is clear that the robust formulation ensures local convergence, 
which also results in smoother changes of topologies as the anisotropy orientation $\beta$ varies from $0\degree$ to $90\degree$; the generated holes tend to deform more consistently and more aligned with the anisotropy angle as opposed to the not robust topologies which transform more abruptly. 
We still observe narrow bands of solid material in the topology generated with the input $(\alpha_1, \alpha_2, \beta) = (4,1.0,15\degree)$; 
in this case, the stopping criterion of maximum 50 iterations stops the first optimisation loop prematurely where more iterations are needed to reach convergence. Releasing this condition solves the problem.

\cref{fig_11_b} demonstrates the performance of selected optimised topologies generated by the robust formulation via inputs $(\alpha_1,\alpha_2)=(12,1.0)$ from \cref{fig_11_a}. 
The overall performance trend for the robust topologies is similar to that of the not robust ones. However, as observed above, ensuring local convergence provides more consistency in the results as the anisotropy angle changes. This consistency can be noticed especially in the identification error plot, where the optimised topology for each anisotropy angle is almost always the best to identify the orthotropic material with that anisotropy angle.

\section{Conclusions and outlook}
\label{conclusions}
We proposed a topology optimisation framework for the optimal design of the test specimen to be utilised for the one-shot identification or discovery of constitutive material models. To this end, we employed the density-based topology optimisation approach with a cost function targeting the robustness of the unknown material parameters against the noise in the experimental strain field. The developed framework was then used to generate optimised specimen designs to be tested in uniaxial or biaxial tensile experiments in order to calibrate isotropic as well as orthotropic material parameters in linear elasticity. The main findings can be summarised as follows:

\begin{itemize}
    
    \item The equation system leading to the unknown material parameters involves a matrix $\ten{A}^\mathrm{eqb}$ which encodes all the information on the specimen geometry and boundary conditions. The conditioning of $\ten{A}^\mathrm{eqb}$dictates the robustness of the identified material parameters against noise in the experimental strain data. In order to use a gradient-based optimisation technique, we propose as possible cost function a $p$-norm condition number or the inverse of the matrix determinant, which 
    are continuously differentiable and  independent of the ground-truth material parameters.
    The investigation of these and other cost functions reveals that $1/\det{\!(\ten{A}^\mathrm{eqb})}$ behaves most stably, yields global mesh convergence, generates smoother topologies with fewer artefacts, and leads to the reduction of the identification error. Due to the non-convex nature of the cost function, the initial guess has a significant impact on the optimised topologies; 
    a multi-start approach with various random initialisations potentially leads to the best design.
    
    \item The global convergence (i.e., mesh independence) of the optimised topologies is achieved through consistent definitions of the weighting factors $\lambda_r$ and $\lambda_q$ appearing in $\ten{A}^\mathrm{eqb}$. 
    Local convergence (i.e., ensuring the existence of a minimum length scale) in the optimised topologies is attained by implementing the robust formulation to topology optimisation, involving projection filtering with three different threshold values. 
    
    \item For the efficient handling of a large number of design variables in topology optimisation, the sensitivities are calculated analytically via the adjoint method, which requires solving one additional FE-like problem. Moreover, the use of the \mbox{PDE} filter rather than the convolution-type operation reduces the memory consumption remarkably. %Furthermore, computations were done in a highly vectorised format and benefited from efficient MATLAB built-in functionalities such as the \textit{sparse} operator to maximise the performance.
    
    \item The effect of input material parameters on the optimised topologies is investigated in the context of linear elastic orthotropic materials. Due to the normalised cost function, the anisotropy descriptors $\alpha_1 = E_{xx}/E_{yy}$, $\alpha_2 = G_{xy}/G_{xy}^{sv}$ and, most importantly, $\beta$ (i.e., the anisotropy orientation) influence the optimised topologies whereas the individual values of the stiffness components are not relevant. 
    
    \item The optimised topologies were evaluated in calibrating orthotropic material parameters. Due to the different relevance of the stiffness components under different anisotropy orientations $\beta$, the hardest parameter to calibrate in a uniaxial tensile test for $\beta<45\degree$ is $E_{xx}$, for $\beta>45\degree$ is $E_{yy}$, and for $\beta=45\degree$ is $G_{xy}$. The topologies optimised with different anisotropy orientations perform almost equally well, indicating a low influence of the input anisotropy orientation. Performance checks against reference topologies confirm the good performance of the optimised designs with few but optimally designed holes.
\end{itemize}

As future outlook, the performance of the optimised topologies in the calibration of constitutive parameters is to be experimentally evaluated. The current framework can be extended to more complex constitutive behaviours, e.g. hyperelasticity or dissipative behaviour. The further inclusion of \mbox{DIC} metrological features could lead to more practically optimised specimens.

\section*{Acknowledgement}
SG and LDL acknowledge funding from the Swiss National Science Foundation through grant N. 200021\_204316 ``Unsupervised data-driven discovery of material laws". Also, they thank Prof. Fabrice Pierron for fruitful discussion. SG thanks student Felix J. Lupp for his effort in improving the efficiency of the code.

\section*{Declaration of competing interest}
The authors declare that they have no known competing financial interests or personal relationships that could have appeared to influence the work reported in this paper.

\section*{Code availability}
The MATLAB code is publicly available at \url{https://gitlab.ethz.ch/compmech/euclid-top}.

\bibliographystyle{elsarticle-harv}
\bibliography{References}

\appendix
\section{\texorpdfstring{Consistent definition of the weighting parameters $\lambda_r$ and $\lambda_q$}{Consistent definition of the weighting parameters lambda\_r and lambda\_q}}
\label{app_weighting_factor}

\setcounter{figure}{0} \renewcommand{\thefigure}{\Alph{section}.\arabic{figure}}
\setcounter{table}{0} \renewcommand{\thetable}{\Alph{section}.\arabic{table}}
\setcounter{subsection}{0} \renewcommand{\thesubsection}{\Alph{section}.\arabic{subsection}}

As introduced in \cref{eq_cost16a}, there exist two weighting parameters $\lambda_r$ and $\lambda_q$ in the definition of $\ten{A}^\mathrm{eqb}$. Here, we provide insight into the consistent definition of these parameters such that the cost functions defined in \cref{eq_topopt05,eq_topopt06} behave mesh convergent. Note that here we do not perform any topology optimisation, but rather look at a fixed sample geometry, i.e., the plate with one hole, as shown in \cref{fig_02_a}. The elastic anisotropic material considered here has $6$ independent stiffness components gathered in $\ten{\theta} = [323,100.03,50.015,190,80.024,144.930]^T~\mathrm{GPa}$.

\begin{figure}[htb]
    \centering
    \begin{subfigure}{0.9\textwidth}
        \centering
        \includegraphics[width=\linewidth]{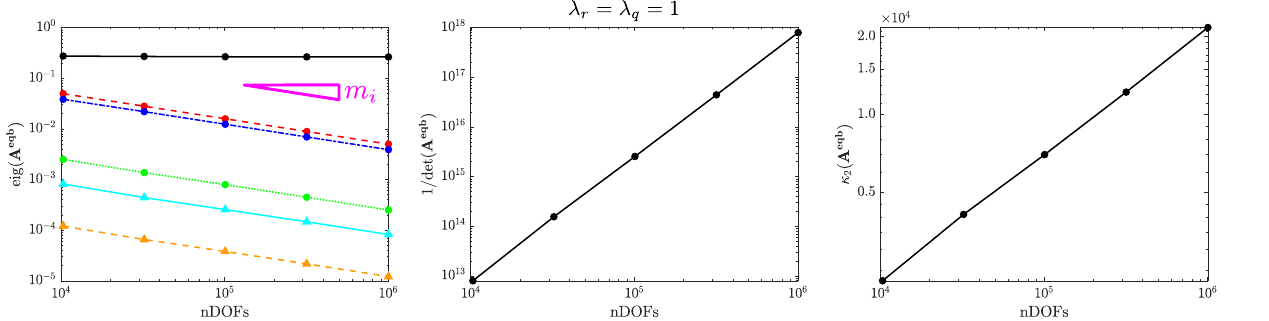} \caption{} \label{fig_app01_a}
    \end{subfigure}
    \begin{subfigure}{0.3\textwidth}
        \centering
        \includegraphics[width=\linewidth]{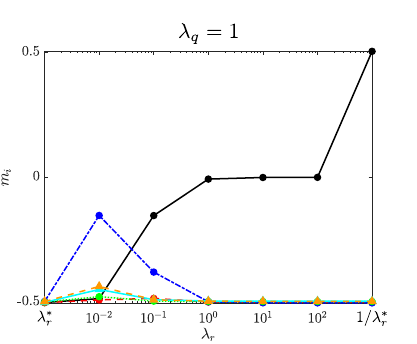} \caption{} \label{fig_app01_b}
    \end{subfigure}
    \begin{subfigure}{0.3\textwidth}
        \centering
        \includegraphics[width=\linewidth]{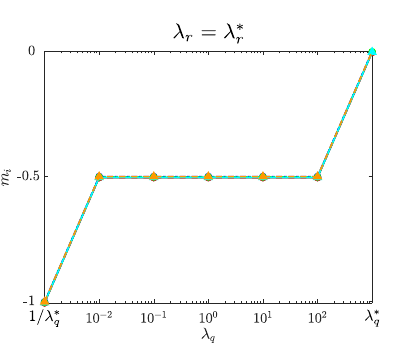} \caption{} \label{fig_app01_c}
    \end{subfigure}
    \begin{subfigure}{0.9\textwidth}
        \centering
        \includegraphics[width=\linewidth]{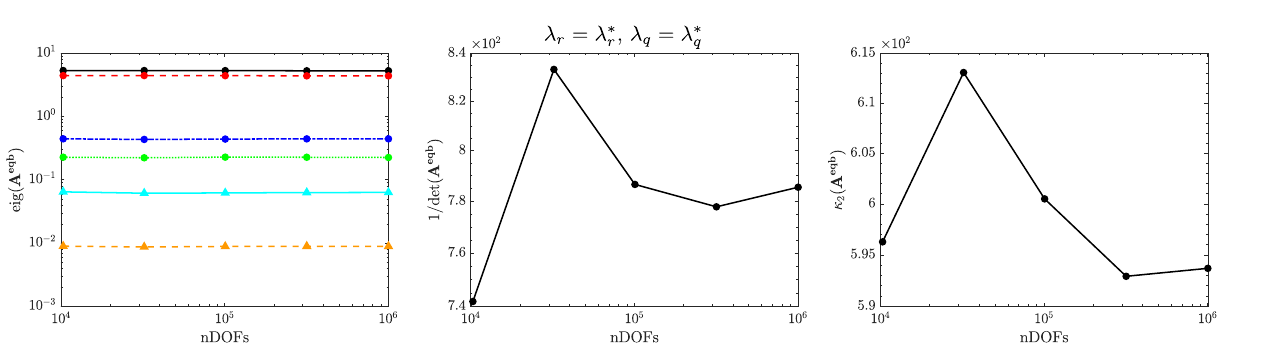} \caption{} \label{fig_app01_d}
    \end{subfigure}
    \caption[]{Consistent definition of the weighting parameters $\lambda_r$ and $\lambda_q$: (a) starting with $\lambda_r=\lambda_q=1$ and plotting $\mathrm{eig}(\ten{A}^\mathrm{eqb})$ (sorted, each with a slope $m_i$), $1/\det(\!\ten{A}^\mathrm{eqb})$ and $\kappa_2(\ten{A}^\mathrm{eqb})$ versus \mbox{nDOFs}; (b) fixing $\lambda_q=1$ and finding $\lambda_r=\lambda_r^{*}$ to unify the slopes $m_i$; (c) fixing $\lambda_r=\lambda_r^{*}$ and finding $\lambda_q=\lambda_q^{*}$ such that the unified slopes $m_i$ approach zero; (d) consistent values $\lambda_r=\lambda_r^{*}$ and $\lambda_q=\lambda_q^{*}$ lead to mesh-convergent behaviour. The eigenvalues and their corresponding slopes are in $mm^{2}$, the determinant is in $mm^{2n_f}$, and the condition number is dimensionless.}
    \label{fig_app01}
\end{figure}

\cref{fig_app01_a} visualises, in logarithmic scale, the eigenvalues, the inverse of the determinant and the 2-norm condition number of $\ten{A}^\mathrm{eqb}$ against \mbox{nDOFs} as the FE mesh is refined. The eigenvalues, sorted in descending order, give $n_f$ lines each with a slope $m_i \, (i=1,\ldots, n_f)$. Hence, $1/\det(\!\ten{A}^\mathrm{eqb})$ gives a line with the slope $-\sum_{i=1}^{n_f} {m_i}$, and $\kappa_2(\ten{A}^\mathrm{eqb})$ gives a line with the slope $(m_1 - m_{n_f})$. In \cref{fig_app01_a}, $\lambda_r=\lambda_q=1$, $m_1 \approx 0$ and $m_i \, (i=2,\ldots,n_f) \approx -0.5$. We aim at finding $\lambda_r$ and $\lambda_q$ such that all $m_i \approx 0$, thereby inducing mesh convergence in $1/\det(\!\ten{A}^\mathrm{eqb})$ and $\kappa_2(\ten{A}^\mathrm{eqb})$.

Let us first look at $\lambda_r$. This weighting parameter is required to balance the contributions of the free and fixed DOFs in the system of linear equations in \cref{eq_cost17} since $\abs{\mathcal{D}^\mathrm{free}}$ is usually significantly greater than $\sum_{s=1}^{n_s}{\abs{\mathcal{D}^{\mathrm{fix},s}}}$. To find the consistent value for $\lambda_r$ denoted here as $\lambda_r^{*}$, we initially fix $\lambda_q=1$ and plot the slopes $m_i$ for different $\lambda_r$ values, see \cref{fig_app01_b}. The line slopes $m_i$ acquire various values while $\lambda_r$ changes, but there is a value of $\lambda_r,\lambda_r^{*}$, at which all slopes are equal, i.e., $m_i \approx -0.5$:
\begin{equation} \label{eq_app_a_01}
    \lambda_r^{*} =\dfrac{\sum_{s=1}^{n_s}{\abs{\mathcal{D}^{\mathrm{fix},s}}}}{\abs{\mathcal{D}^\mathrm{free}}} \, .
\end{equation}
\noindent
Note that with this setting the slope of $\kappa_2(\ten{A}^\mathrm{eqb})$ already reduces to zero, whereas $1/\det(\!\ten{A}^\mathrm{eqb})$ monotonically increases at the rate $n_f/2$. This calls for the need to define the parameter $\lambda_q$. 

The weighting factor $\lambda_q$ can help tailor the unified $m_i$ value at zero (rather than $-0.5$). \cref{fig_app01_c} sheds light on the influence of $\lambda_q$ on the slopes $m_i$ (while $\lambda_r=\lambda_r^{*}$) and therefore on the convergence behaviour of $1/\det(\!\ten{A}^\mathrm{eqb})$. As evidenced in this plot, there exists a $\lambda_q=\lambda_q^{*}$ at which $m_i \approx 0$:
\begin{equation} \label{eq_app_a_02}
    \lambda_q^{*} =\sqrt{\sum_{s=1}^{n_s}{\abs{\mathcal{D}^{\mathrm{fix},s}}} + \abs{\mathcal{D}^\mathrm{free}}} \, .
\end{equation}

Upon the consistent setting of the weighting parameters, the plots from \cref{fig_app01_a} are redrawn in \cref{fig_app01_d}. It is observed that all the mentioned quantities have become independent of the FE mesh, hence, $1/\det(\!\ten{A}^\mathrm{eqb})$ and $\kappa_2(\ten{A}^\mathrm{eqb})$ can be used as cost functions in the context of topology optimisation.

\section{Other explored cost functions}
\label{app_other_cost}

\setcounter{figure}{0} \renewcommand{\thefigure}{\Alph{section}.\arabic{figure}}
\setcounter{table}{0} \renewcommand{\thetable}{\Alph{section}.\arabic{table}}
\setcounter{subsection}{0} \renewcommand{\thesubsection}{\Alph{section}.\arabic{subsection}}

This section discusses alternative cost definitions (explored for automated specimen design) and their pros and cons.

\subsection{Minimum distance between strain data points}
\label{app_min_dist}

One way to formulate the specimen design problem is by looking at the principal strain space and its coverage by the data points, i.e., strains at the Gauss points. The more widely and uniformly spread the data points, the richer the deformation field experienced by the specimen, and so, the more accurate the identified material parameters. In this context, a cost function based on the minimum distance between the strain data points can be defined as:
\begin{equation} \label{eq_app_b_01}
    \mathrm{cost}_{\mathrm{alt},2} = -\left(\sum_{i=1}^{n(n-1)/2}{\left(\dfrac{d_i}{\mathrm{min}\!\left(\ten{d}_{@ \mathrm{init.}}\right)} \right)^{p}} \right)^{\frac{1}{p}} \, ,
\end{equation}
\noindent
where $\ten{d} \in \mathbb{R}^{n(n-1)/2}$ represents the vector containing all the unique pairwise Euclidean distances between $n$ points in the principal strain space, and $\mathrm{min}\!\left(\ten{d}_{@ \mathrm{init.}}\right)$ denotes the minimum distance between the points at initialisation. The $\mathcal{l}_p$-regularisation term is used instead of the $\mathrm{min}(\cdot)$ operator to enable analytical differentiation.

Topology optimisation based on the cost in \cref{eq_app_b_01} aims at maximising the minimum distance between the data points, thus providing extensive coverage over the principal strain space. The advantage of such a cost definition lies in the design of a specimen topology encompassing various deformation modes as it undergoes a simple uniaxial tensile test. On the other hand, the main downside of such a cost is its intrinsic \textit{locality}, which only targets the minimum distance between the strain data points (i.e., a local feature), therefore lacking a \textit{global} view of the whole design optimisation problem. Moreover, the minimisation of \cref{eq_app_b_01} would necessarily require a random field for initialisation with $\mathrm{min}\!\left(\ten{d}_{@ \mathrm{init.}}\right) > 0$, meaning that evenly distributed design densities (inducing concentrated and overlapping strain points) would be impractical. 
\subsection{Frobenius-norm condition number}
\label{app_fro_norm_cond}

An alternative to the 2-norm condition number is the condition number based on the Frobenius norm:
\begin{equation} \label{eq_app_b_03}
    \kappa_F{\left(\ten{A}^\mathrm{eqb}\right)} = {\norm{{\ten{A}^\mathrm{eqb}}^{\phantom{l\!}}}}_F \,{\norm{{\ten{A}^\mathrm{eqb}}^{-1}}}_F \, ,
\end{equation}
\noindent
where the Frobenius norm for ${\ten{A}^\mathrm{eqb}}$ with vectorised components $A_i^\mathrm{eqb}$ ($i=1,\ldots,n_f^2$) is given by
\begin{equation} \label{eq_topopt03}
    {\norm{\ten{A}^\mathrm{eqb}}}_F = \left(\sum_{i=1}^{n_f^2} {\abs{A_{i}^\mathrm{eqb}}}^2 \right)^{\frac{1}{2}} \, .
\end{equation}

It can be shown that the Frobenius norm of a matrix is always greater than or equal to its $2$-norm (utilising Cauchy–Schwarz inequality). Thus the following bound relation can be found:

\begin{equation} \label{eq_app_b_05}
    1 \leq \kappa_2{\left(\ten{A}^\mathrm{eqb}\right)} = {\norm{{\ten{A}^\mathrm{eqb}}^{\phantom{l\!}}}}_2 \, {\norm{{\ten{A}^\mathrm{eqb}}^{-1}}}_2 \leq {\norm{{\ten{A}^\mathrm{eqb}}^{\phantom{l\!}}}}_F \,{\norm{{\ten{A}^\mathrm{eqb}}^{-1}}}_F = \kappa_F{\left(\ten{A}^\mathrm{eqb}\right)} \, . 
\end{equation}
\noindent
One can thus define the cost for topology optimisation based on $\kappa_F(\ten{A}^\mathrm{eqb})$ as follows:
\begin{equation} \label{eq_app_b_06}
    \mathrm{cost}_{\mathrm{alt},3} = \dfrac{\kappa_F{\left(\ten{A}^\mathrm{eqb}\right)}}{\kappa_F{\left(\ten{A}_{@ \mathrm{init.}}^\mathrm{eqb}\right)}} \, ,
\end{equation}
\noindent
where $\kappa_F(\ten{A}_{@ \mathrm{init.}}^\mathrm{eqb})$ denotes the Frobenius-norm condition number at initialisation.
 Even though $\kappa_F$ is insensitive to the strain levels, we still normalise it by the corresponding value at the initial configuration; this is found to help increase the stability of the optimisation process (as pointed out earlier in \cref{topopt}). 
Unfortunately, the implementation of \cref{eq_app_b_06} in the optimisation framework leads to unstable behaviour of the optimiser and abrupt changes of the topology during the course of optimisation, to the extent that, in some cases, the output topology from one machine can vary slightly from the one of another machine. For this reason, we do not further use this cost definition.
\subsection{\texorpdfstring{$p$-norm condition number}{p-norm condition number}}
\label{app_p_norm_cond}

\begin{figure}[tb]
    \centering
    \begin{subfigure}{\textwidth}
        \centering
        \includegraphics[width=1\linewidth]{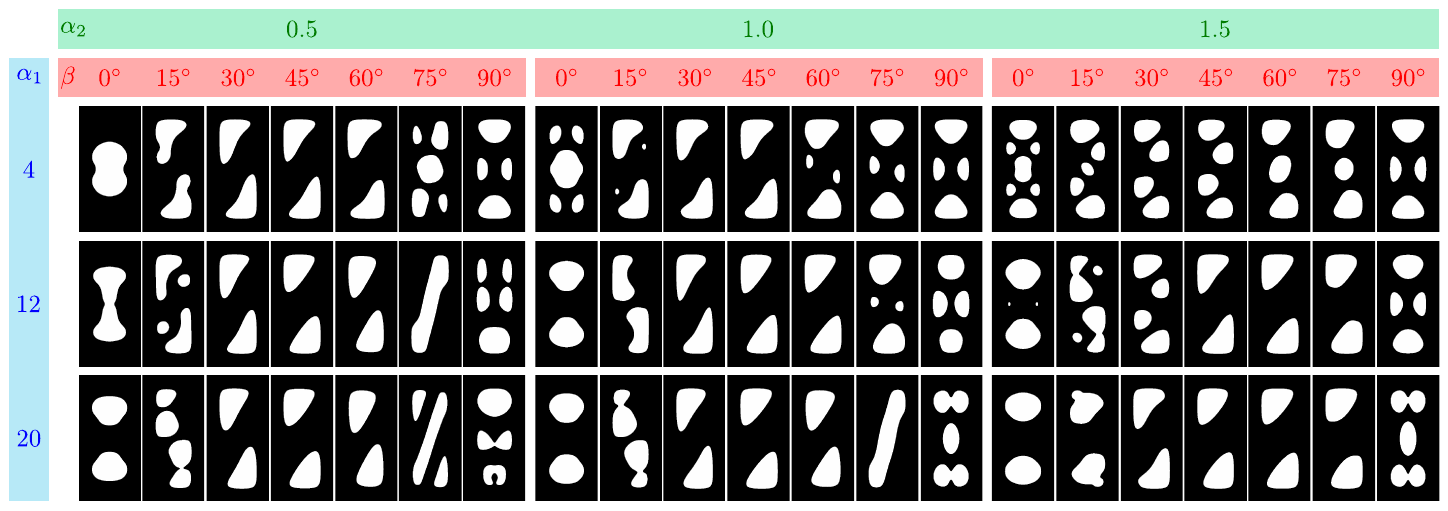} \caption{} \label{fig_app02_a}
    \end{subfigure}
    \begin{subfigure}{\textwidth}
        \centering
        \includegraphics[width=1\linewidth]{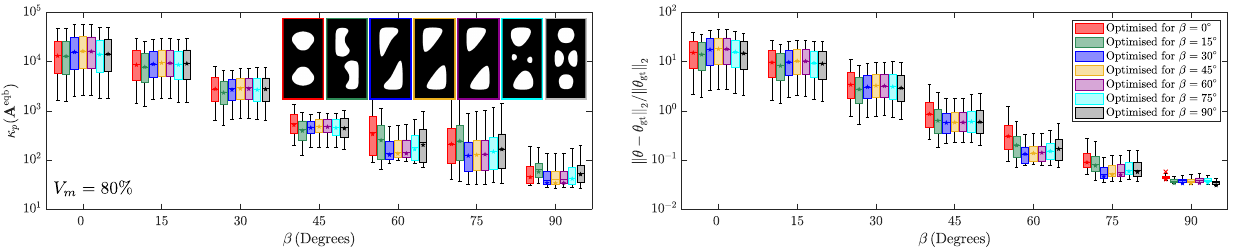} \caption{} \label{fig_app02_b}
    \end{subfigure}
    \caption[]{The $p$-norm condition number as cost function: (a) optimised topologies with $p=8$ for different orthotropic materials. The densities are initialised evenly, and the material volume constraint is set to $V_m = 80\%$; (b) identification of the orthotropic material parameters using the optimised topologies obtained with $(\alpha_1,\alpha_2)=(12,1.0)$ from part (a). The left plot visualises the $p$-norm condition number (dimensionless) while the right one demonstrates the identification error (in percentage), both presented against the anisotropy angle $\beta$.}
    \label{fig_app02}
\end{figure}

The $p$-norm condition number in \cref{eq_topopt01} is based on the matrix $p$-norm in \cref{eq_topopt02}. 
To establish a bound relation with respect to the Frobenius norm (and hence to the $2$-norm), we leverage H\"older's inequality for vectors $\ten{a}$ and $\ten{b}$, with $r>1$:
\begin{equation} \label{eq_app_b_09}
    \sum_{i=1}^{n} \abs{a_i} \abs{b_i} \leq \left(\sum_{i=1}^{n}{ {\abs{a_i}}^{r}}\right)^{\frac{1}{r}} \left(\sum_{i=1}^{n}{ {\abs{b_i}}^{\frac{r}{r-1}}}\right)^{1-\frac{1}{r}} , \quad \forall \quad \ten{a},\ten{b} \in \mathbb{R}^n \text{ or } \mathbb{C}^n \, , r>1  \, .
\end{equation}
\noindent
By substituting $\abs{a_i} = {\abs{A_i^\mathrm{eqb}}}^2$, $\abs{b_i} = 1$ and $r = p/2>1$, and taking the square root, we obtain
\begin{equation} \label{eq_app_b_11}
    \left(\sum_{i=1}^{n_f^2} {\abs{A_i^\mathrm{eqb}}}^2\right)^\frac{1}{2} = {\norm{{\ten{A}^\mathrm{eqb}}}}_F \leq n_f^{1-\frac{2}{p}} \left(\sum_{i=1}^{n_f^2}{ \abs{A_i^\mathrm{eqb}}^{p}}\right)^{\frac{1}{p}} = n_f^{1-\frac{2}{p}} {\norm{\ten{A}^\mathrm{eqb}}}_p  \, .
\end{equation}
\noindent
Such a bound relation can be obtained similarly for $(\ten{A}^\mathrm{eqb})^{-1}$, giving the final inequality:
\begin{equation} \label{eq_app_b_12}
    1 \leq \kappa_2{\left(\ten{A}^\mathrm{eqb}\right)} \leq \kappa_F{\left(\ten{A}^\mathrm{eqb}\right)} = {\norm{{\ten{A}^\mathrm{eqb}}^{\phantom{l\!}}}}_F \, {\norm{{\ten{A}^\mathrm{eqb}}^{-1}}}_F \leq n_f^{2\left(1-\frac{2}{p}\right)} {\norm{{\ten{A}^\mathrm{eqb}}^{\phantom{l\!}}}}_p \, {\norm{{\ten{A}^\mathrm{eqb}}^{-1}}}_p = n_f^{2\left(1-\frac{2}{p}\right)} \kappa_p{\left(\ten{A}^\mathrm{eqb}\right)}  \, . 
\end{equation}
\noindent
Therefore, the product $n_f^{2(1-\frac{2}{p})} \kappa_p{(\ten{A}^\mathrm{eqb})}$ is an upper bound for $\kappa_F{(\ten{A}^\mathrm{eqb})}$ which in turn is an upper bound for $\kappa_2{(\ten{A}^\mathrm{eqb})}$.

The results for an isotropic material using $\kappa_8$ in \cref{eq_topopt04} were already given in \cref{fig_05}. Next, \cref{fig_app02_a} presents the results for orthotropic elasticity using the input material parameters in \cref{tab_results01}. The gallery of topologies is generated with the material volume fraction $V_m = 80\%$. The optimised topologies are rather similar to those obtained with the original determinant-based cost function (shown in \cref{fig_07}). However, a few of the $\kappa_8$-optimised topologies violate the special type of symmetry observed earlier in \cref{fig_07}, see e.g.~those obtained with $(\alpha_1,\alpha_2,\beta) = (12,0.5,90\degree)$, $(20,0.5,15\degree)$ and $(20,0.5,90\degree)$. Moreover, narrow-width topological features are more frequent here, see e.g.~$(\alpha_1,\alpha_2,\beta) = (4,1.0,15\degree)$, $(12,1.5,0\degree)$ and $(20,1.0,15\degree)$. The performance of selected topologies generated via inputs $(\alpha_1,\alpha_2)=(12,1.0)$ from the gallery is demonstrated in \cref{fig_app02_b}. The plots of cost and identification error visualise similar trends to their counterparts in \cref{fig_08}, where the topology obtained with $\beta = 15\degree$ is superior in terms of cost and identification error for many anisotropy orientations.

In view of the above, the cost function based on $\kappa_p$ is not superior to the determinant-based cost function, hence, we do not further use it here.
\section{Density filtering}
\label{filter}
\setcounter{figure}{0} \renewcommand{\thefigure}{\Alph{section}.\arabic{figure}}
\setcounter{table}{0} \renewcommand{\thetable}{\Alph{section}.\arabic{table}}
\setcounter{subsection}{0} \renewcommand{\thesubsection}{\Alph{section}.\arabic{subsection}}

\begin{figure}[tb]
    \centering
    \includegraphics[width=0.65\linewidth]{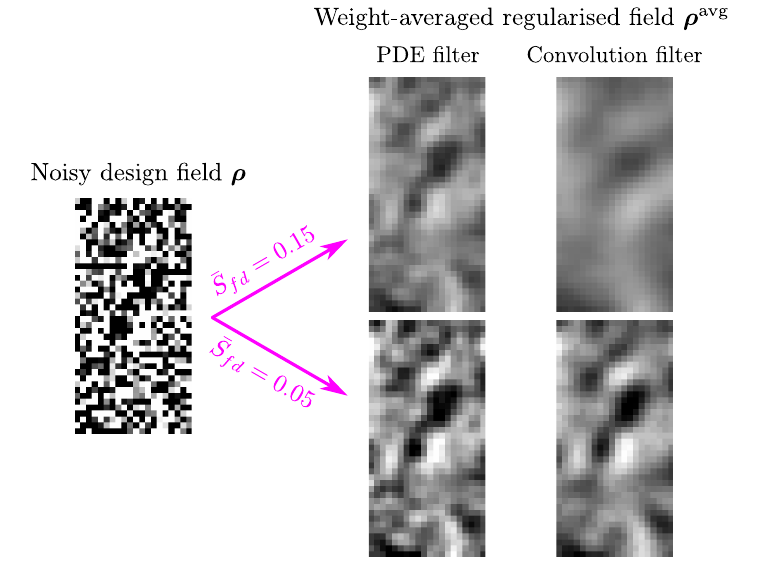}
    \caption[]{Weight-averaged density filtering to regularise the topology optimisation problem: comparing the implicit \mbox{PDE} and explicit convolution-type filters for two different neighbourhood sizes defined by $\bar{S}_{fd}$ over a noisy design field.}
    \label{fig_03}
\end{figure}

As briefly discussed in \cref{density_based_topology_optimisation}, topology optimisation without regularisation is an ill-posed problem. For this reason, filtering techniques have been developed. Among the different choices proposed in the literature, a popular filtering technique is the so-called weight-averaged, density filtering over a neighbourhood of radius $r_\mathrm{min}$ \citep{Bruns2001, Bourdin2001} which acts upon the design densities $\rho_e$ as a convolution-type operator and transforms them into the averaged densities
\begin{equation} \label{eq_topopt09}
    \rho_e^\mathrm{avg} = \dfrac{1}{\sum_{i=1}^{n_e}{H_{e,i}}} \sum_{i=1}^{n_e}{H_{e,i} \rho_i} \, ,
\end{equation}
with the convolution kernel 
\begin{equation} \label{eq_topopt09b}
    H_{e,i} = \max{\!\left( 0, r_\mathrm{min} - \Delta{\left(e,i\right)} \right)} \, .
\end{equation}
\noindent
Here, $\Delta{\left(e,i\right)}$ denotes the distance between the centres of two elements $e$ and $i$, which dictates how much the density of element $i$ (within the neighbourhood $r_\mathrm{min}$) contributes to the averaged density of element $e$. The filtering radius $r_\mathrm{min}$ is defined as 
\begin{equation} \label{eq_topopt10}
    r_\mathrm{min} = \dfrac{r_\mathrm{min}^\mathrm{abs}}{L_e} \, ,
\end{equation}
with $r_\mathrm{min}^\mathrm{abs}$ as the absolute filtering radius and $L_e$ as the element size. The ratio between the area of the absolute filtering neighbourhood $\pi (r_\mathrm{min}^\mathrm{abs})^2$ and the domain area %(with dimensions $L_X$ and $L_Y = \bar{a}_r \, L_X$
is kept constant and equal to %$0.15$ 
$\bar{S}_{fd} = 0.15$ 
(see \cref{fig_02_b}).
This setup leads to a mesh-independent filtering effect.
The filtering radius $r_\mathrm{min}$ dictates the minimum size of topological features allowed to form (in both material and void regions). There exists, however, a trade-off in setting $r_\mathrm{min}$ (or equivalently $\bar{S}_{fd}$): in general, increasing $r_\mathrm{min}$ results in broader averaging and stronger regularisation, hence larger topological features but also more grey scales; conversely, reducing $r_\mathrm{min}$ gives rise to the formation of tiny features and checkerboard patterns with less grey scales in the optimised topology (see \cref{fig_03}). 
Therefore, a \textit{sweet spot} can be found (most often with trial and error) which renders sufficiently large features and not many grey scales.

The convolution kernel can be pre-computed, assembled into a sparse matrix $\ten{H} \in \mathbb{R}^{n_e \times n_e}$, and then used throughout the optimisation process for matrix-vector multiplication with the vector of design densities. Here a twofold problem arises: first, when the filtering radius $r_\mathrm{min}$ is increased for a bigger neighbourhood search (while keeping $n_e$ constant), the computational complexity and the memory utilisation to compute $\ten{H}$ grows quadratically for 2D problems; second, the number of nonzero entries in $\ten{H}$ grows significantly (first linearly, and then almost logarithmically until saturation) which hinders the benefits of the sparse matrix-vector multiplication of $\ten{H}$ with $\ten{\rho}$.
To solve the memory issue explained above,
we resort to the implicit filter proposed by \cite{Lazarov2011}. The density filtering in \cref{eq_topopt09} is implicitly expressed by the solution of a Helmholtz-type \mbox{PDE} with homogeneous Neumann boundary conditions:
\begin{equation} \label{eq_topopt11}
    \left\{ \begin{array}{{l}}  
    {-R_\mathrm{min}^2 \nabla^2{\!\rho^\mathrm{avg}} + \rho^\mathrm{avg} = \rho} \, , \\[0.75ex]
    {\dfrac{\partial{\rho^\mathrm{avg}}}{\partial{\ten{n}}}} = 0 \, ,
    \end{array} \right.
\end{equation}
\noindent
where $\rho^\mathrm{avg}$ and $\rho$ are the continuous representations of the averaged and design densities, respectively, and $\ten{n} \in \mathbb{R}^{2}$ is the outward normal vector to the boundary. The parameter $R_\mathrm{min}$ acts similarly to $r_\mathrm{min}$, with an approximate relation between them derived by \cite{Lazarov2011}:
\begin{equation} \label{eq_topopt12}
    R_\mathrm{min} = \dfrac{r_\mathrm{min}}{2\sqrt{3}} \, .
\end{equation}
\noindent
\cref{fig_03} illustrates the fairly similar behaviours of the \mbox{PDE} and convolution-type filters when they are applied to the same input noisy design field. The convolution-type filter regularises the topology more strongly, especially when $\bar{S}_{fd}$ is relatively large. This is rooted in the definition of the approximate relation in \cref{eq_topopt12}.

As put forward by \cite{Lazarov2011}, the FE discretisation of the \mbox{PDE} in \cref{eq_topopt11} can be formulated as a linear system with the solution $\ten{\rho}^\mathrm{avg} \in \mathbb{R}^{n_e}$ as the weight-averaged densities: 
\begin{equation} \label{eq_topopt13}
    \ten{K}_\mathrm{F} \left(\ten{T}_\mathrm{F} \, \ten{\rho}^\mathrm{avg} \right) = \left( \ten{T}_\mathrm{F} \, \ten{\rho} \right) \, ,
\end{equation}
\noindent
where $\ten{K}_\mathrm{F} \in \mathbb{R}^{\abs{\mathcal{D}_\mathrm{F}} \times \abs{\mathcal{D}_\mathrm{F}}}$ with $\mathcal{D}_\mathrm{F}=\{(a,1): a=1,\ldots,n_n\}$ is the standard FE stiffness matrix for scalar problems (including the factor $R_\mathrm{min}^2$ in its definition), and $\ten{T}_\mathrm{F} \in \mathbb{R}^{\abs{\mathcal{D}_\mathrm{F}} \times n_e}$ is a matrix that maps the element density values to a vector of respective nodal values. With this approach, the cost of filtering and the size of the involved matrices become independent of the filtering radius, which helps with memory usage reduction. However, the use of the \mbox{PDE} filter requires the solution of the additional FE problem in \cref{eq_topopt13}, which, however, can be achieved efficiently by one-time computation and storage of the factorisation of $\ten{K}_\mathrm{F}$ (refer to \cite{Andreassen2011} for implementation details).

As discussed earlier, density filtering produces intermediate density values, graphically represented as grey scales. Even though this is necessary for problem regularisation and convergence, grey scales are undesirable in the output topology since they cannot be manufactured in practice. To convert the grey-scaled into a black-and-white topology, techniques such as projection filtering have been introduced \citep{Guest2004,Sigmund2007,Xu2010}, see \cite{Wang2011} for a review. These projection filters can be applied on top of the density averaging schemes.
In the simplest projection, if $\rho_e^\mathrm{avg} \geq \phi$ (with $\phi$ as a chosen 
 threshold) then the projected density, denoted as $\rho_e^\mathrm{phys}$, becomes $\rho_e^\mathrm{phys} = 1$, and if $\rho_e^\mathrm{avg} < \phi$ then $\rho_e^\mathrm{phys} = 0$. %\cite{Wang2011} reviews some of the conventional threshold projection methods, among which 
The projection given by 
\begin{equation} \label{eq_topopt14}
    \ten{\rho}^\mathrm{phys} = \dfrac{\tanh{\!\left(\psi \phi \right)} + \tanh{\!\left(\psi \left(\ten{\rho}^\mathrm{avg} - \phi \right) \right)}}{\tanh{\!\left(\psi \phi \right)} + \tanh{\!\left(\psi \left(1 - \phi \right) \right)}}
\end{equation}
\noindent
is a computationally efficient choice as it encodes the above threshold conditions in a vectorised expression.
Here, $\psi$ is a regularisation parameter. It is customary to start the optimisation process with $\psi= 2^{0} = 1$ (i.e., almost no projection) 
to let the topology evolve freely, and to increase $\psi$ gradually as the optimisation goes on.
\cite{Andreassen2011} suggest to gradually double $\psi$ with respect to the previous value whenever the stopping criterion is met until reaching $\bar{\psi} = 2^{9} = 512$,
which approximately creates a step function that outputs binary designs. The stopping criterion is introduced in \cref{algorithm}. 
\cref{fig_04} gives the flowchart of the optimisation algorithm where the continuation scheme for increasing $\psi$ forms the outer loop. The threshold parameter $\phi$ in \cref{eq_topopt14} is set to 0.5.

\section{Analytical sensitivity calculation}
\label{app_sensitivity}

\setcounter{figure}{0} \renewcommand{\thefigure}{\Alph{section}.\arabic{figure}}
\setcounter{table}{0} \renewcommand{\thetable}{\Alph{section}.\arabic{table}}
\setcounter{subsection}{0} \renewcommand{\thesubsection}{\Alph{section}.\arabic{subsection}}

Due to the employment of a gradient-based optimisation approach, we need to derive the analytical gradients (i.e., sensitivities) of the cost function with respect to the design variables $\ten{\rho}$. Recalling the chain rule in \cref{eq_topopt15}, we find the three major derivatives here, namely $\dfrac{d{\left(\mathrm{cost}\right)}}{d{\ten{\rho}^\mathrm{phys}}} \in \mathbb{R}^{1 \times n_e}$, $\dfrac{d{\ten{\rho}^\mathrm{phys}}}{d{\ten{\rho}^\mathrm{avg}}} \in \mathbb{R}^{n_e \times n_e}$ and $\dfrac{d{\ten{\rho}^\mathrm{avg}}}{d{\ten{\rho}}} \in \mathbb{R}^{n_e \times n_e}$:

\begin{enumerate}[label=(\Roman*)]
\item Derivation of $\dfrac{d{\left(\mathrm{cost}\right)}}{d{\ten{\rho}^\mathrm{phys}}}$

The sensitivity of the cost function in \cref{eq_topopt05} with respect to the physical densities $\ten{\rho}^\mathrm{phys}$ can be found from:
\begin{equation} \label{eq_app_c_01}
    \dfrac{d{\left(\mathrm{cost}\right)}}{d{\ten{\rho}^\mathrm{phys}}} = \dfrac{\partial \left(\mathrm{cost}\right)}{\partial \ten{\rho}^\mathrm{phys}} + \dfrac{\partial \left(\mathrm{cost}\right)}{\partial \ten{U}} \dfrac{d{\ten{U}}}{d{\ten{\rho}^\mathrm{phys}}}  \,.
\end{equation}
\noindent
We start with the first term
\begin{equation} \label{eq_app_c_02}
    \dfrac{\partial \left(\mathrm{cost}\right)}{\partial \ten{\rho}^\mathrm{phys}} = \dfrac{d{\left(\mathrm{cost}\right)}}{d{\ten{A}_\mathrm{glob}}} \dfrac{\partial \ten{A}_\mathrm{glob}}{\partial \ten{\rho}^\mathrm{phys}} \, ,
\end{equation}
\noindent
where
\begin{subequations} \label{eq_app_c_03}
\begin{gather}
    \dfrac{d{\left(\mathrm{cost}\right)}}{d{\ten{A}_\mathrm{glob}}} = \dfrac{d{\left(\mathrm{cost}\right)}}{d{\ten{A}^\mathrm{eqb} }} \dfrac{d{\ten{A}^\mathrm{eqb} }}{d{\ten{A} }} \dfrac{d{\ten{A}}}{d{\ten{A}_\mathrm{glob}}} \, , \label{eq_app_c_03a} \\
    \dfrac{d{\left(\mathrm{cost}\right)}}{d{\ten{A}^\mathrm{eqb} }} = -\dfrac{\det{\!\left(\ten{A}_{@ \mathrm{init.}}^\mathrm{eqb}\right)}}{\det{\!\left(\ten{A}^\mathrm{eqb}\right)}} \underbrace{\left(\ten{A}^\mathrm{eqb}\right)^{-1}}_{\mapsto (1 \times {n_f}^2)}   \, , \label{eq_app_c_03b} \\
     \dfrac{d{\ten{A}^\mathrm{eqb} }}{d{\ten{A} }} = \left(\underbrace{\left(\dfrac{d{\ten{A}^T}}{d{\ten{A}}}\right)^T \ten{A}}_{\mapsto ( {n_f}(\abs{\mathcal{D}^\mathrm{free}} + n_s) \times {n_f}^2 )}\right)^T + \underbrace{\ten{A}^T \left(\dfrac{d{\ten{A}}}{d{\ten{A}}}\right)}_{\mapsto ({n_f}^2 \times {n_f}(\abs{\mathcal{D}^\mathrm{free}} + n_s))}  \, , \label{eq_app_c_03c} \\
     \dfrac{d{\ten{A}}}{d{\ten{A}_\mathrm{glob}}} = \dfrac{d{\ten{A}}}{d{\ten{A}_\mathrm{glob}^\mathrm{free}}} \dfrac{d{\ten{A}_\mathrm{glob}^\mathrm{free}}}{d{\ten{A}_\mathrm{glob}}} + \dfrac{d{\ten{A}}}{d{\ten{A}_\mathrm{glob}^\mathrm{fix}}} \dfrac{d{\ten{A}_\mathrm{glob}^\mathrm{fix}}}{d{\ten{A}_\mathrm{glob}}} \, , \label{eq_app_c_03d}
\end{gather}
\end{subequations}
\noindent
and 
\begin{equation} \label{eq_app_c_04}
     \dfrac{\partial \ten{A}_\mathrm{glob}}{\partial \ten{\rho}^\mathrm{phys}} = \bigcup_{e=1}^{n_e} {{3\left(\rho_e^\mathrm{phys}\right)^2 \left(1-\rho_\mathrm{min} \right)} \, \underbrace{\ten{A}_e{\left(\ten{U}_e \right)}}_{\mapsto (\abs{\mathcal{D}_e}{n_f} \times 1)}} \, . 
\end{equation}
\noindent
In the equations above, we have introduced the operator ${\mapsto (d_x \times d_y)}$ to denote the reshaping of an array to $\mathbb{R}^{d_x \times d_y}$. In \cref{eq_app_c_03c,eq_app_c_03d} there are some remaining derivatives in the form of sparse constant matrices defined only once (at the first iteration) using the \textit{sparse} operator in MATLAB, stored in the memory, and then used directly during the optimisation process. For brevity, we do not write these derivatives here.

Next, we compute the second term.
For this, we first derive the state (i.e., equilibrium) equation in \cref{eq_topopt06b} to find $\dfrac{d{\ten{U}}}{d{\ten{\rho}^\mathrm{phys}}} 
\in \mathbb{R}^{\abs{\mathcal{D}} \times n_e}$ as follows
\begin{equation} 
     \dfrac{d{\ten{K}}}{d{\ten{\rho}^\mathrm{phys}}} \ten{U} + \ten{K} \dfrac{d{\ten{U}}}{d{\ten{\rho}^\mathrm{phys}}} = \ten{0} \Rightarrow \dfrac{d{\ten{U}}}{d{\ten{\rho}^\mathrm{phys}}} = - \ten{K}^{-1} \left(\dfrac{d{\ten{K}}}{d{\ten{\rho}^\mathrm{phys}}} \ten{U} \right) \, . \label{eq_app_c_05a}
\end{equation}
This derivation requires solving a linear system with the right-hand side as the large matrix $\dfrac{d{\ten{K}}}{d{\ten{\rho}^\mathrm{phys}}} \ten{U} \in \mathbb{R}^{\abs{\mathcal{D}} \times n_e}$ given by
\begin{equation}
     \dfrac{d{\ten{K}}}{d{\ten{\rho}^\mathrm{phys}}} \ten{U} = \bigcup_{e=1}^{n_e} {{3\left(\rho_e^\mathrm{phys}\right)^2 \left(1-\rho_\mathrm{min} \right)} \, \ten{K}_e{\left(\ten{\theta}\right)} \,\ten{U}_e} \, . \label{eq_app_c_05b} 
\end{equation}

In order to avoid solving $n_e$ systems of linear equations in \cref{eq_app_c_05a}, which is computationally expensive, we exploit the so-called \textit{adjoint} method (i.e., backward derivation) \citep{Johnson2012}. Hence, we look for the product $\dfrac{\partial \left(\mathrm{cost}\right)}{\partial \ten{U}} \dfrac{d{\ten{U}}}{d{\ten{\rho}^\mathrm{phys}}}$, given by 
\begin{equation}
    \dfrac{\partial \left(\mathrm{cost}\right)}{\partial \ten{U}} \dfrac{d{\ten{U}}}{d{\ten{\rho}^\mathrm{phys}}} = -\underbrace{\dfrac{\partial \left(\mathrm{cost}\right)}{\partial \ten{U}} \ten{K}^{-1}}_{\ten{\gamma}^{T}}  \left( \dfrac{d{\ten{K}}}{d{\ten{\rho}^\mathrm{phys}}} \ten{U} \right) = - \ten{\gamma}^{T} \left( \dfrac{d{\ten{K}}}{d{\ten{\rho}^\mathrm{phys}}} \ten{U} \right) \, , \label{eq_app_c_06a}
\end{equation}

where we have introduced the adjoint vector $\ten{\gamma} \in \mathbb{R}^{\abs{\mathcal{D}}}$ which is the solution to only \textit{one} system of linear equations, i.e., the adjoint state equation
\begin{equation}
    \ten{K} \ten{\gamma} = \left( \dfrac{\partial \left(\mathrm{cost}\right)}{\partial \ten{U}} \right)^{T} \, . \label{eq_app_c_06b}
\end{equation}

The right-hand side of the adjoint system of equations is constructed by $\left( \dfrac{\partial \left(\mathrm{cost}\right)}{\partial \ten{U}} \right)^{T} \in \mathbb{R}^{\abs{\mathcal{D}}}$ given by:
\begin{subequations} \label{eq_app_c_07}
\begin{gather}
    \dfrac{\partial \left(\mathrm{cost}\right)}{\partial \ten{U}} = \dfrac{d{\left(\mathrm{cost}\right)}}{d{\ten{A}_\mathrm{glob}}} \dfrac{\partial \ten{A}_\mathrm{glob}}{\partial \ten{U}}  \, , \label{eq_app_c_07a} \\
    \dfrac{\partial \ten{A}_\mathrm{glob}}{\partial \ten{U}} = \bigcup_{e=1}^{n_e} {\tilde{\rho}_e^\mathrm{phys} \, \underbrace{\dfrac{d{\ten{A}_e{\left(\ten{U}_e \right)}}}{d{\ten{U}_e}}}_{\mapsto (\abs{\mathcal{D}_e}{n_f} \times \abs{\mathcal{D}_e})}} \, , \label{eq_app_c_07b} \\
    \dfrac{d{\ten{A}_e{\left(\ten{U}_e \right)}}}{d{\ten{U}_e}} = \dfrac{d{\ten{A}_e}}{d{\tilde{\ten{\varepsilon}}_e^\mathcal{h}}}  \dfrac{d{\tilde{\ten{\varepsilon}}_e^\mathcal{h}}}{d{\hat{\ten{\varepsilon}}_e^\mathcal{h}}} \dfrac{d{\hat{\ten{\varepsilon}}_e^\mathcal{h}}}{d{\ten{U}_e}} \approx \sum_{i=1}^{n_\mathrm{gp}} {\sum_{j=1}^{n_\mathrm{gp}} {\ten{B}_e^T{\! \left(\xi_i,\eta_j \right)} \underbrace{\dfrac{d{\tilde{\ten{\varepsilon}}_e^\mathcal{h}}}{d{\hat{\ten{\varepsilon}}_e^\mathcal{h}}} \ten{B}_e{ \left(\xi_i,\eta_j \right)}}_{\mapsto (3 \times {n_f}\abs{\mathcal{D}_e})} \det{\!\left(\ten{J}_e{ \left(\xi_i,\eta_j \right)} \right)} \times w_i \times w_j  }} \, , \label{eq_app_c_07c} \\
    \dfrac{d{\tilde{\ten{\varepsilon}}_e^\mathcal{h}}}{d{\hat{\ten{\varepsilon}}_e^\mathcal{h}}} = 
    \left[\! \! \begin{array}{*{18}{c}}
            1 & 0 & 0 & 0 & 1 & 0 & 0 & 0 & 1 & 0 & 0 & 0 & 0 & 0 & 0 & 0 & 0 & 0 \\
            0 & 0 & 0 & 1 & 0 & 0 & 0 & 0 & 0 & 0 & 1 & 0 & 0 & 0 & 1 & 0 & 0 & 0 \\
            0 & 0 & 0 & 0 & 0 & 0 & 1 & 0 & 0 & 0 & 0 & 0 & 0 & 1 & 0 & 0 & 0 & 1 \\
            \end{array} \! \!\right]^{T} \, . \label{eq_app_c_07d}
\end{gather}
\end{subequations}

The adjoint method is beneficial since it enables the efficient computation of the sensitivities at the cost of solving only one extra linear system (whose cost is comparable to that of one FE analysis) rather than solving $n_e$ linear systems.

\item Derivation of $\dfrac{d{\ten{\rho}^\mathrm{phys}}}{d{\ten{\rho}^\mathrm{avg}}}$

The projection filter in \cref{eq_topopt14} can be derived to obtain:
\begin{equation} \label{eq_app_c_08}
    \dfrac{d{\ten{\rho}^\mathrm{phys}}}{d{\ten{\rho}^\mathrm{avg}}} = \mathrm{diag}{\left(\dfrac{\psi \, \mathrm{sech}^2{\left(\psi \left(\ten{\rho}^\mathrm{avg} - \phi \right) \right)}}{\tanh{\!\left(\psi \phi \right)} + \tanh{\!\left(\psi \left(1 - \phi \right) \right)}}\right)} \,.
\end{equation}
\noindent
The operator $\mathrm{diag}(\cdot)$ creates a square diagonal matrix whose diagonal components are the entries of the input vector.

\item Derivation of $\dfrac{d{\ten{\rho}^\mathrm{avg}}}{d{\ten{\rho}}}$

Following the expression of the \mbox{PDE} filter in \cref{eq_topopt13}, it is possible to find:
\begin{equation} \label{eq_app_c_09}
    \dfrac{d{\ten{\rho}^\mathrm{avg}}}{d{\ten{\rho}}} = \ten{T}_\mathrm{F}^{T} \ten{K}_\mathrm{F}^{-1} \left( \ten{T}_\mathrm{F} \, \ten{I}_{n_e} \right) \,,
\end{equation}
\noindent
where $\ten{I}_{n_e} \in \mathbb{R}^{n_e \times n_e}$ is the identity matrix.
\end{enumerate}

Note that $\dfrac{d{\left(\mathrm{cost}\right)}}{d{\ten{\rho}}}$ can be computed in a more efficient manner \citep{Andreassen2011}:
\begin{equation} \label{eq_app_c_10}
    \dfrac{d{\left(\mathrm{cost}\right)}}{d{\ten{\rho}}} = \left( \ten{T}_\mathrm{F}^{T} \ten{K}_\mathrm{F}^{-1} \left( \ten{T}_\mathrm{F} \, \left( \left( \dfrac{d{\left(\mathrm{cost}\right)}}{d{\ten{\rho}^\mathrm{phys}}} \right)^{T} \cdot \left( \dfrac{\psi \, \mathrm{sech}^2{\left(\psi \left(\ten{\rho}^\mathrm{avg} - \phi \right) \right)}}{\tanh{\!\left(\psi \phi \right)} + \tanh{\!\left(\psi \left(1 - \phi \right) \right)}} \right)  \right) \right) \right)^{T} \, ,
\end{equation}
\noindent
with $\cdot$ denoting the element-wise product of two vectors resulting in a vector with the same size.

The analytical computation of sensitivities would be more involved for more complex material behaviours (e.g., geometric/material non-linearities, dissipation). In such cases, one can resort to automatic differentiation. However, the principles of the adjoint method remain the same.

\section{The effect of initialisation in the optimised topologies}
\label{app_init}

\setcounter{figure}{0} \renewcommand{\thefigure}{\Alph{section}.\arabic{figure}}
\setcounter{table}{0} \renewcommand{\thetable}{\Alph{section}.\arabic{table}}
\setcounter{subsection}{0} \renewcommand{\thesubsection}{\Alph{section}.\arabic{subsection}}

As pointed out in \cref{filter}, our cost definition for topology optimisation is a non-convex and non-linear function of the design variables $\ten{\rho}$. The non-convexity of the cost gives rise to the existence of numerous local minima in the high-dimensional space, which implies that the initial guess may significantly influence the reached solution, i.e., the obtained optimised design. In \cref{orthotropic} we noticed that the topology optimised for a given anisotropy orientation $\beta$ was not always the best for the identification of orthotropic materials with the same value of $\beta$. To dig further into this issue, we briefly examine the influence of the initial density field on the results of our topology optimisation framework.

\begin{figure}[tb]
    \centering
    \begin{subfigure}{\textwidth}
        \centering
        \includegraphics[width=\linewidth]{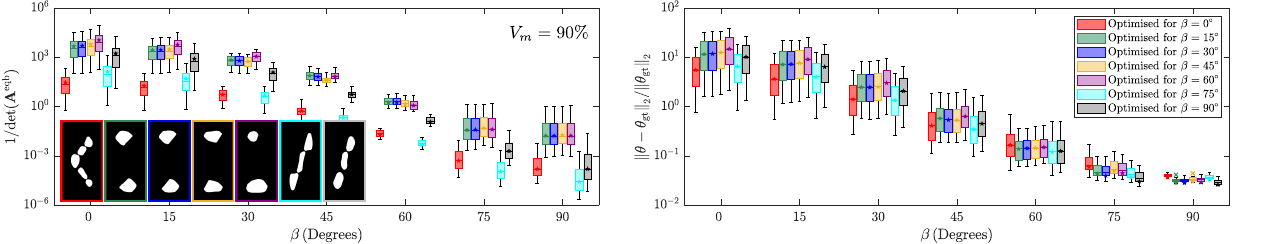} \caption{} \label{fig_app03_a}
    \end{subfigure}
    \begin{subfigure}{\textwidth}
        \centering
        \includegraphics[width=\linewidth]{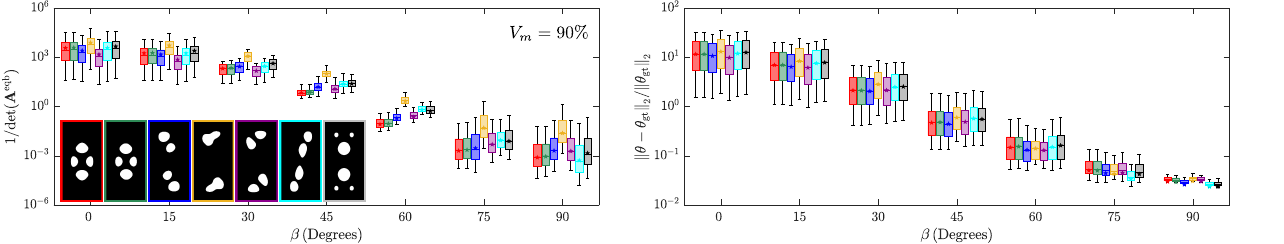} \caption{} \label{fig_app03_b}
    \end{subfigure}
    \caption[]{The effect of initialisation on the optimised topologies, the unnormalised cost (in $mm^{-12}$) and the identification error (in percentage) (shown only for $V_m=90\%$ for brevity): (a) initialisation with $20\%$ noise in design densities, illustrated in \cref{fig_02_c}; (b) initialisation from the interim result of $\beta=0\degree$, illustrated in \cref{fig_02_d}. These results are comparable to those from initialisation with evenly distributed densities in \cref{fig_08}(c).}
    \label{fig_app03}
\end{figure}

\cref{fig_app03} illustrates the effect of initialisation in the optimised topologies and their performance in the identification of orthotropic materials. For conciseness, the results of this analysis are reported only for the material volume fraction $V_m = 90\%$ which led to the lowest identification error bounds among the different volume fractions shown in \cref{fig_08}. Let us first look into the effect of noisy (rather than even) initialisations. To this aim, we generate noisy fields by perturbing the even density distribution with Gaussian noise with zero mean and $20\%$ standard deviation, a sample of which is plotted in \cref{fig_02_c}. Further, in a multi-start setup, we create $50$ different noise patterns (i.e., seeds), and feed them into the topology optimisation framework. For each anisotropy orientation $\beta$, the topology with the lowest (unnormalised) cost is selected as the optimal topology. \cref{fig_app03_a} visualises the optimised topologies obtained from the prescribed noisy initial fields together with their performance assessment graphs. As evident, the overall looks of the topologies are quite similar to the respective results from the evenly distributed initialisation (owing to the regularisation by the weight-averaged density filter), except for the topologies generated with $\beta=0\degree$ and $90\degree$. Noisy-initiated optimised topologies, however, do not exhibit the special type of symmetry\footnote{Noisy fields with standard deviations less than $15\%$ are practically smoothened out with the application of the weight-averaged density filtering, hence, they often produce symmetric results resembling those obtained with the even initialisation.} originally observed in \cref{fig_07}. Additionally, due to the introduction of noise, there exists a tendency for the formation of holes close to each other, which, as explained earlier, lowers the cost and the identification error. Also in these plots (and even to a larger extent than with even initialisation) we observe that optimised topologies with $\beta = 0\degree$ and $75\degree$ lead to the best results almost in all cases. 
Possibly, following the procedure above but multi-starting with a much larger number of seeds (say 500), we could arrive at the expected situation where each $\beta$-optimised topology is the best for the identification of materials with that value of $\beta$. We did not try this. Otherwise, the robust topology optimisation approach showed promise to address this issue (see \cref{results_robust_topopt}).

\cref{fig_app03_b} plots the results achieved by initialising the design field with the densities obtained after the first loop of optimisation (see \cref{fig_02_d}) for the inputs $(\alpha_1,\alpha_2,\beta) = (12,1.0,0\degree)$ as an example. 
Since the "interim" densities are symmetric, the optimised topologies also turn out to be symmetric. We notice a quite different appearance of the topologies for almost all $\beta$ angles, highlighting the crucial impact of the initial guess. Because of the significantly modified topologies, the costs and identification errors also perform quite differently than in the previous cases. Noteworthy is that no unique superior topology can be detected in this case. 

\section{Implementation aspects in MATLAB and computational performance}
\label{comp_perform}

\setcounter{figure}{0} \renewcommand{\thefigure}{\Alph{section}.\arabic{figure}}
\setcounter{table}{0} \renewcommand{\thetable}{\Alph{section}.\arabic{table}}
\setcounter{subsection}{0} \renewcommand{\thesubsection}{\Alph{section}.\arabic{subsection}}

In this section, we report some details on the implementation aspects and the computational performance of the developed algorithm. %for the automatic design of the specimen topology for constitutive law identification.
Generally, topology optimisation aims at minimising a cost function dependent on a large number of design variables (in the order of $10^5-10^6$) through an iterative algorithm. Thus, the implementation must be efficient to ensure reasonable performance. In a MATLAB environment, it is essential that the cost and the sensitivities are computed in a highly vectorised fashion, thereby avoiding slow \textit{for}-loops. When it comes to creating sparse matrices and performing assembly operations efficiently, the \textit{sparse} operator in MATLAB is essential \citep{Andreassen2011}. The syntax reads $\ten{S} = \textit{sparse}(\ten{i},\ten{j},\ten{v},d_x,d_y)$ and generates the matrix $\ten{S} \in \mathbb{R}^{d_x \times d_y}$ from the triplets $\ten{i}$, $\ten{j}$, $\ten{v}$ such that $\ten{S}(\ten{i}(k),\ten{j}(k)) = \ten{v}(k)$ for any entry $k$ \citep{MATLAB_Sparse}. Another implementation aspect is the use of the \textit{memoize} command to cache the outputs of expensive functions and return the cached value (instead of its re-evaluation) given the function is called with the same inputs \citep{MATLAB_Memoize}. This functionality can be applied to the filtering operation since the filtered densities are required three times per iteration, namely, to compute the cost and the constraint and to plot the evolving topology. This is inevitable since sharing the filtered densities at each iteration of the optimisation loop among different functions is not straightforward in MATLAB.

To gain more efficiency in the optimisation framework, many constant FE- and cost-related calculations (partly thanks to the fixed structured mesh) need to be executed only once at the beginning and then stored in the memory as persistent variables to be readily available throughout the optimisation process. The same practice can be followed to precompute and cache large constant sparse matrices as well as the position vectors $\ten{i}$ and $\ten{j}$ for variable-valued sparse matrices. Note that position vectors can be stored as single-precision arrays to decrease the memory footprint. As pointed out in \ref{app_sensitivity}, to efficiently compute the sensitivities of the cost with respect to the design variables, the adjoint method should be employed which requires one additional FE-like system of equations instead of $n_e$. Also, with the use of the \mbox{PDE} filter rather than the convolution-type operation, the memory usage is reduced substantially since the problem size and complexity of the former do not depend on the filtering radius.

Building on the implementation aspects discussed above, our topology optimisation algorithm for optimal specimen design with $157922$ design variables (i.e., the converged discretisation level) runs in about 6 hours on a desktop machine with Core i7 $3.8~\mathrm{GHz}$ processor, requiring only $8~\text{GB}$ of random access memory throughout the computations. When using the robust formulation, the wall-clock time roughly reaches 14.5 hours, whereas parallelisation over three processors speeds up the algorithm by nearly 24\% resulting in a run time of approximately 11 hours. Investigating the optimisation process, it is revealed that the MATLAB optimiser engine \textit{fmincon} sometimes requires too many cost evaluations to solve the saddle-point problem (i.e., the Karush-Kuhn-Tucker conditions) and to compute the search direction in each iteration \citep{Rojas-Labanda2015}. To improve the temporal performance, it might be helpful to try other available optimiser engines, which require fewer function evaluations per iteration, e.g. the method of moving asymptotes \citep{Svanberg1987}. 

\end{document}